\documentclass[manuscript]{aastex}

\usepackage{epsfig}

\slugcomment{}

\shorttitle{}
\shortauthors{Centeno}

\begin{document}


\title{The naked emergence of solar active regions observed with SDO/HMI}


\author{Rebecca Centeno}
\affil{High Altitude Observatory (NCAR), 3080 Center Green Dr., Boulder CO 80301}

\begin{abstract}
We take advantage of the HMI/SDO instrument to study the naked emergence of active regions from the first imprints of the magnetic field on the solar surface.
To this end, we followed the first 24 hours in the life of two rather isolated ARs that appeared on the surface when they were about to cross the central meridian.
We analyze the correlations between Doppler velocities and the orientation of the vector magnetic field finding, consistently, that the horizontal fields connecting the main polarities are dragged to the surface by relatively-strong upflows and are associated to elongated granulation
that is, on average, brighter than its surroundings. The main magnetic footpoints, on the other hand, are dominated by vertical fields and downflowing plasma. The appearance of
moving dipolar features, MDFs, (of opposite polarity to that of the AR) in between the main footpoints, is a rather common occurrence once the AR reaches a certain size. The buoyancy  of the fields is insufficient to lift up the magnetic arcade as a whole. Instead, weighted by the plasma that it carries, the field is pinned down to the photosphere at several places in between the main footpoints, giving life to the MDFs and enabling channels of downflowing plasma. MDF poles tend to drift towards each other, merge and disappear. This is likely to be the signature of a reconnection process in the dipped field lines, which relieves some of the weight allowing the magnetic arcade to finally rise beyond the detection layer of the HMI spectral line.

\end{abstract}


\section{Active region emergence}

Solar active regions have puzzled humanity for millenia. The work of
\citet{hale1908} established that sunspots harbour strong magnetic
fields, which, together with other observations of the properties of
the solar cycle, set the basis for the current theories of solar dynamo.
Although specific details of the emergence of active regions (AR) are still a matter of
research, the established paradigm says that their magnetic field is generated close to the base of the convection
zone. Then, presumably triggered by deep convective flows and buoyant
instabilities, these magnetic fields rise towards the surface, protrude through it
and leave footprints in the form of sunspots and plage \citep[see, for
instance, the reviews by][]{fmi1997,yuhong2009}.

Observationally, active region emergence sites have been the object of
many studies. Most of these works, especially the ones from the earlier days,
made use of intensity and circular polarization measurements, which
restricted the inference of the magnetic field vector to its
longitudinal component only (i.e. projected along the line-of-sight, LOS).  

\noindent A series of papers \citep{1985zwaan, 1985brants_a, 1985brants_b, brants_steenbeek} marked the starting point 
for the analysis of the magnetic and dynamic properties of the
emergence process. Based on the width and Doppler shifts of the Stokes I profiles, they inferred
$\sim 500{\rm G}$ transverse magnetic fields associated to small-scale
upflows in the photosphere, which they interpreted to be the signatures
of the top of the flux loops reaching the surface. They observed strong downflows in
the vicinity of rapidly growing pores, but no systematic flows within
the pores themselves. They also reported the existence of
very strong (1000-2000G) transverse fields, result which was later
questioned by \citet{lites1998} on the basis of the limitations in the
determination of the transverse component of the magnetic field using only
intensity and circular polarization measurements.
\citet{strous_thesis} and \citet{strous1999} saw elongated darkenings in the photospheric continuum
intensity at emergence sites associated to $0.5{\rm kms^{-1}}$ upflows. These darkenings
were almost aligned with the orientation of the AR, which lead the authors to 
interpret them as being the crests of undulatory flux tubes penetrating the photosphere.

\citet{lites1998} used full Stokes measurements of small emerging
bipolar regions to determine the properties of the magnetic field
vector. They found horizontal fields of 200-600 Gauss associated to
small transient upflows at the site of the emergence, suggesting that
there is a canopy of weak horizontal magnetic fields over-arching the
emergence zone. Small flux elements constantly drift away from the emergence site and
attain kiloGauss strengths only when they become almost vertically
oriented.
\citet{kubo2003} derived magnetic filling factors of around 80\% at
the emergence site, indicating that the horizontal fields of the tops
of the magnetic arcades are intrinsically weak.

\citet{2002bernasconi} reported on small moving dipolar features (MDFs) of opposite
polarity to that of the active region, that appear in the midst of the
emerging sites. They found that these features tended to flow into sunspots and
supergranule boundaries. The authors identified MDFs as stitches where the
emerging flux ropes were still tied to the photosphere by trapped
mass, giving the emerging field a serpentine nature.
\citet{vargas2012} observed small-scale short-lived dark
features followed by brightenings in the low chromosphere associated 
to these serpentine fields. They were interpreted as signatures of the energy release due to
reconnection of U-loops and elementary arch filament systems rising up into the chromosphere. 
\citet{watanabe} suggested that the emergence of flux tubes inside
active regions is a triggering mechanism of Ellerman bombs. They
proposed several scenarios in which the newly emerged magnetic fields
either interact with pre-existing fields or suffer reconnections
within their dipped arcade structures. In both cases, the magnetic
reconnection happens in the chromosphere, releasing the energy
responsible for the observed Ellerman Bombs.  

There are quite a few works that focus on the velocity flows in young
emerging active regions. While it is accepted that systematic velocities in the umbrae of
mature sunspots are insignificant, the flows during their formation
seem to be a matter of debate. From the observational point of view,
there has been a range of qualitatively different results over the
last few decades. Symmetric downflows at the footpoints of individual
emerging loops were reported by \citet{1985brants_b}. However, asymmetries
in the flows have also been reported by several authors. From
the predominant downflows in the leading part of
young active regions found by \citet{walton} and later \citet{cauzzi1996},
to the upwelling velocities in the umbra of a
following sunspot pointed out by \citet{sigwarth1998}, obsevations
in this respect
seem to be non-conclusive.

Numerical simulations of the formation of ARs focus either on the deep
convection zone or the uppermost 10 Mm and the photosphere. This is
due to the huge range of time- and length-scales involved in the
process.
Fully compressible MHD simulations of the emergence of ARs in the last
10-20\,Mm below the photosphere try to explain the rise of magnetic
flux through a highly stratified atmosphere and the subsequent formation of
coherent sunspots of kilo-Gauss strengths. By letting a buoyant,
twisted semi-torus flux tube be kinematically advected into the
computational domain, \citet{cheung2010} have
rather successfully achieved to explain the formation of an AR and
some of the associated observational properties (elongated granules,
mixed polarity patterns in the emergence zone, pore formation and
light bridges). \citet{stein2011SoPh} and \citet{stein2011} studied a complementary situation
where a uniform, untwisted, horizontal field is advected into the
computational domain by convective inflows through a depth of
20\,Mm. In their simulations, a large-scale magnetic loop emerges
through the surface leading to the formation of a bipolar pore-like structure.

There is a tendency for new flux to emerge within - or in the vicinity of - existing
active regions \citep[see, for instance,][]{zhang2012, mactaggart2011, lites2010, zuccarello2008}, which indicates that the underlying toroidal field is
already disturbed and likely to erupt again \citep{vanballegooijen}. A
vast majority of the existing observations of newly emerged magnetic
flux takes place in this scenario. Observations reliant on non-synoptic
instruments are very unlikely to capture the naked emergence of an
AR. In the context of this paper, the attribute {\em naked} refers to flux
emergence that is isolated from and unrelated to, pre-existing
magnetic activity. This particular characteristic makes it possible to
study the properties of emergence independently of the interaction with
large-scale pre-existing fields.
In order to comprehensively understand the properties of these emergence
sites, the evolution of the full vector magnetic field needs to be
measured. 
The Helioseismic and Magnetic Imager (HMI) suits the bill perfectly. 
Unlike any other instrument before, it provides uninterrupted, high
cadence photospheric vector magnetic field and Doppler velocities of
the full solar disk, rendering it possible to study the magnetic and
dynamic properties of active regions from their very first stages of 
emergence, as long they appear on the visible disk.

\noindent In this paper we attempt to compile the main properties of the
``naked'' emergence of active regions as seen by HMI. 
The particularities of the chosen datasets and a comprehensive study
of their advantages and limitations are analyzed in Section
\ref{sec:observations}. In section \ref{sec:evolution}, the evolution of different magnetic
and dynamic aspects observed during the first stages of emergence are
described and discussed, and a consistent scenario is put together in
section \ref{sec:discussion}.

\section{Observations and data reduction}\label{sec:observations}

In this paper we analyze the properties of two relatively isolated active regions that
emerged when they were about to cross the central meridian of the
solar disk. The data were
taken by the vector camera of the HMI instrument
\citep{hmi_a} on board the Solar
Dynamics Observatory \citep[SDO,][]{sdo}, which provides continuous full disk measurements of the Stokes
vector of the photospheric Fe {\sc i} 6173 \AA\ line every $\sim 135$
seconds. Although this is the native cadence, the standard HMI vector
products correspond to averages in 16-minute tapered windows, rendering a final cadence of 12
minutes. HMI is a filter instrument. The full width at half maximum (FWHM) of the HMI filtergrams is
~76 m\AA\ and the spectral line is sampled at six equispaced wavelength positions.
The spatial resolution of the HMI data is of $\sim 1\arcsec$
($0.5\arcsec$ pixel) and the polarimetric sensitivity about $\sim 1.2\cdot 10^{-3}$
times continuum intensity. 

\noindent The standard pipeline calibration and
the spectral line inversion \citep[Very Fast Inversion of the Stokes
Vector (VFISV),][]{vfisv} of 12-minute averages were used to compute
the photospheric vector magnetic field quantities and Doppler
velocities (see section \ref{sec:inversion}).

\begin{table} 
\begin{center}
\begin{tabular}{c|cccc}
       & date \& initial & hemisphere &
       approx. position & average $\mu$ \\
 & emergence time & & & \\
\hline
AR 11105  & Sep 1, 2010 @ 20:00UT & North & 20N 5E & 0.975 \\
AR 11211  & May 7, 2011 @ 23:00UT & South & 13S 10E & 0.982  \\
\end{tabular}
\end{center}
\caption{Details of the two datasets presented in this paper.\label{tab:datasets}}
\end{table}

Details of the datasets can be found in table \ref{tab:datasets}. Two
active regions, AR 11105 and AR 11211, were followed for 24 hours after the first
signatures of emergence were detected. Of course, the 
magnetic signature precedes any darkening in the continuum
intensity, so we consider the beginning of the process to happen when
magnetograms show the first hints of activity, beyond the signals of pre-existing
network patches.
\noindent SDO has been collecting data since April 2010. These particular events were selected from a very wealthy
dataset for two reasons: 1, they were both relatively isolated from
pre-existing active regions and 2, they took place very close to
disk center, where projection effects are minimal
(the average heliocentric angle, $\mu$, of the center of each AR over
the first 24 hours of its life is listed in table \ref{tab:datasets}). Not intending to
carry out a statistical study of the properties of active region
emergence, we do attempt to sample regions with significantly different flux
output. Whilst AR 11105 grew quickly and developed into a full grown
system with sunspots and pores, AR11211 generated a few tiny pores and
decayed within a matter of days, before leaving the visible disk. 

\subsection{The spectral line inversion}\label{sec:inversion}

In order to obtain the magnetic field and Doppler velocities at the
Photosphere, the HMI full Stokes data were processed with the VFISV
spectral line inversion code \citep{vfisv}. This code inverts the
polarized radiative transfer equation assuming that the solar
atmosphere can be represented by the Milne-Eddington (ME) approximation and that the generation of polarized
radiation takes place within the classical Zeeman effect regime.
These assumptions impose stringent constraints on the possible
solutions for the model atmosphere and lead to limitations in the
interpretation of the Stokes profiles. 
A Milne-Eddington model assumes that all of the physical parameters that
describe the atmosphere are constant along the line of
sight, except for the source function, which varies linearly with the
optical depth. Asymmetries of the Stokes profiles
cannot be interpreted in the context of this approximation since no
gradients in the velocity or the magnetic field are allowed.
Although no depth dependence for the inverted parameters is obtained, the
results are representative of the average physical properties of the
atmosphere in the region of formation of the spectral line \citep{westendorp}.

\noindent VFISV operates assuming a magnetic filling factor of unity, $\alpha =
1$, where $\alpha$ is the fraction of the area of the pixel that
contains magnetic field. This means that
each pixel is considered to be fully magnetized rather than composed
of a magnetic and a non-magnetic components. This assumption has a
different effect on the longitudinal and the transverse components of
the retrieved magnetic field but, in general, it
leads to an underestimation of the field strength and a biasing of
the inclinations towards more horizontal configurations \citep[see,
for instance, the discussion in ][]{sanchezalmeida}.

No magnetic field disambiguation was applied to the
datasets. The magnetic field inclinations, $\theta_B$, reported
in this paper refer to the
line-of-sight and the azimuth angles, $\chi_B$, vary between 0 and 180
degrees in the plane perpendicular to the LOS. Because during the time span of the observations presented in
this paper neither of the ARs was far away from disk center (the
heliocentric angle, $\mu$, was always larger than 0.97; see table \ref{tab:datasets}), the two
solutions for the azimuth of the magnetic field should yield
inclinations with respect to the local vertical that do not differ
much with those calculated with respect to the LOS.

Photon noise that propagates thorough the spectral line inversion process
results in uncertainties in the magnetic field parameters derived
by the inversion. The transverse component of the magnetic field
vector is much more sensitive to this than the longitudinal
component. But random noise is not the only contributor to the
uncertainties. There are systematic effects that come from the
instrument and from the inversion algorithm that are more difficult to
characterize. The instrumental effects produce variations in
time, with periods that are related to the orbital motion of the satellite.
If we take a statistically significant sample of pixels whose
polarization profiles are pure noise and we
give them to the inversion code, the resulting average field strength
will vary between 70 and 100\,G, as a function of the orbital velocity of
the satellite and the position on the solar disk. For a given location
on the disk and a given orbital velocity, the spread in the
derived field strength is of $\pm30\,{\rm G}$. To be on the
conservative side we adopted a noise level of 150\,G for the total
field strength and at 100\,G for its longitudinal component, ${\rm B_{LOS}}$.

\subsection{Velocity calibration}

\begin{figure}
\begin{center}
\includegraphics[angle=0, scale=0.3]{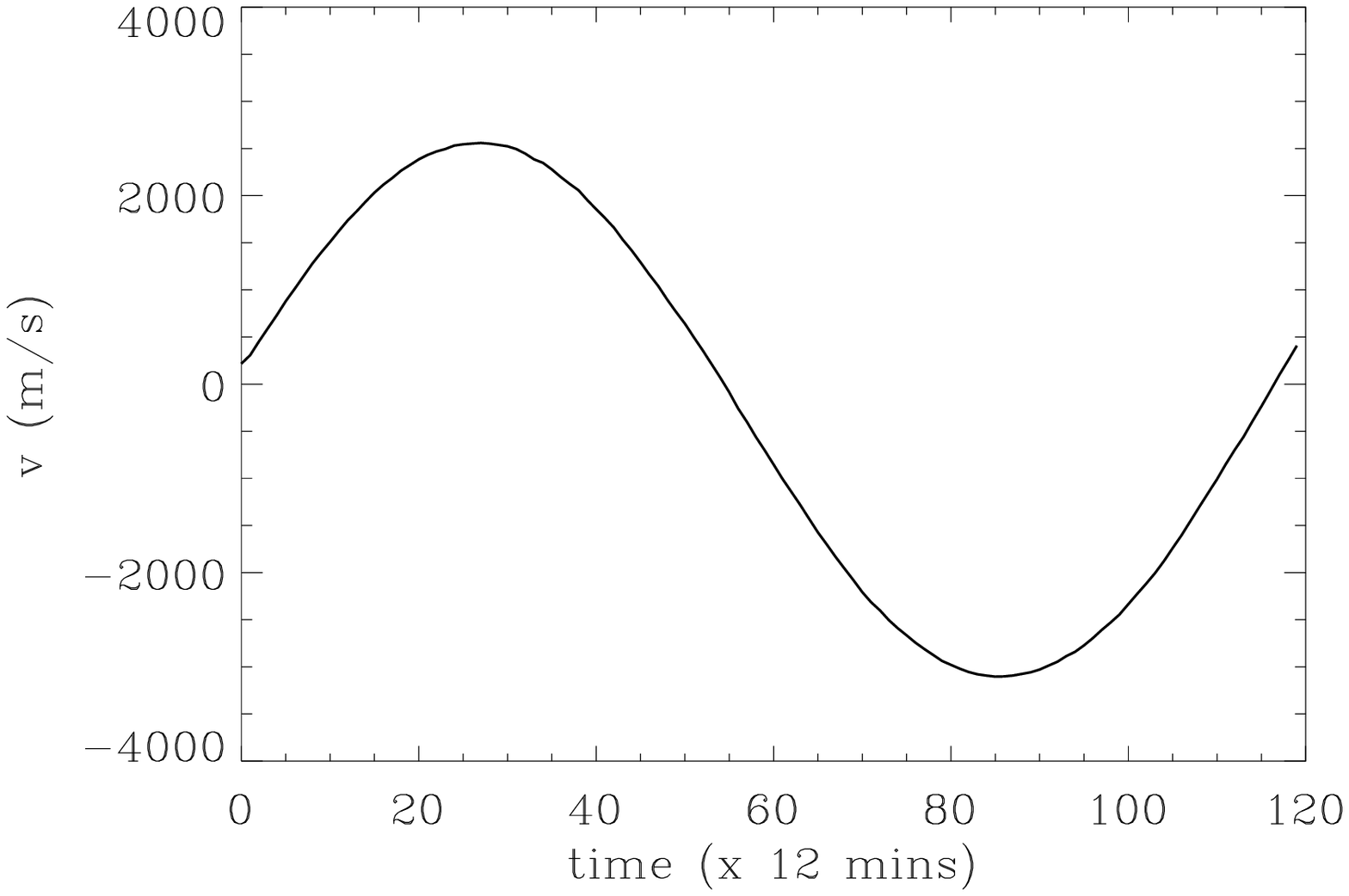}
\includegraphics[angle=0, scale=0.3]{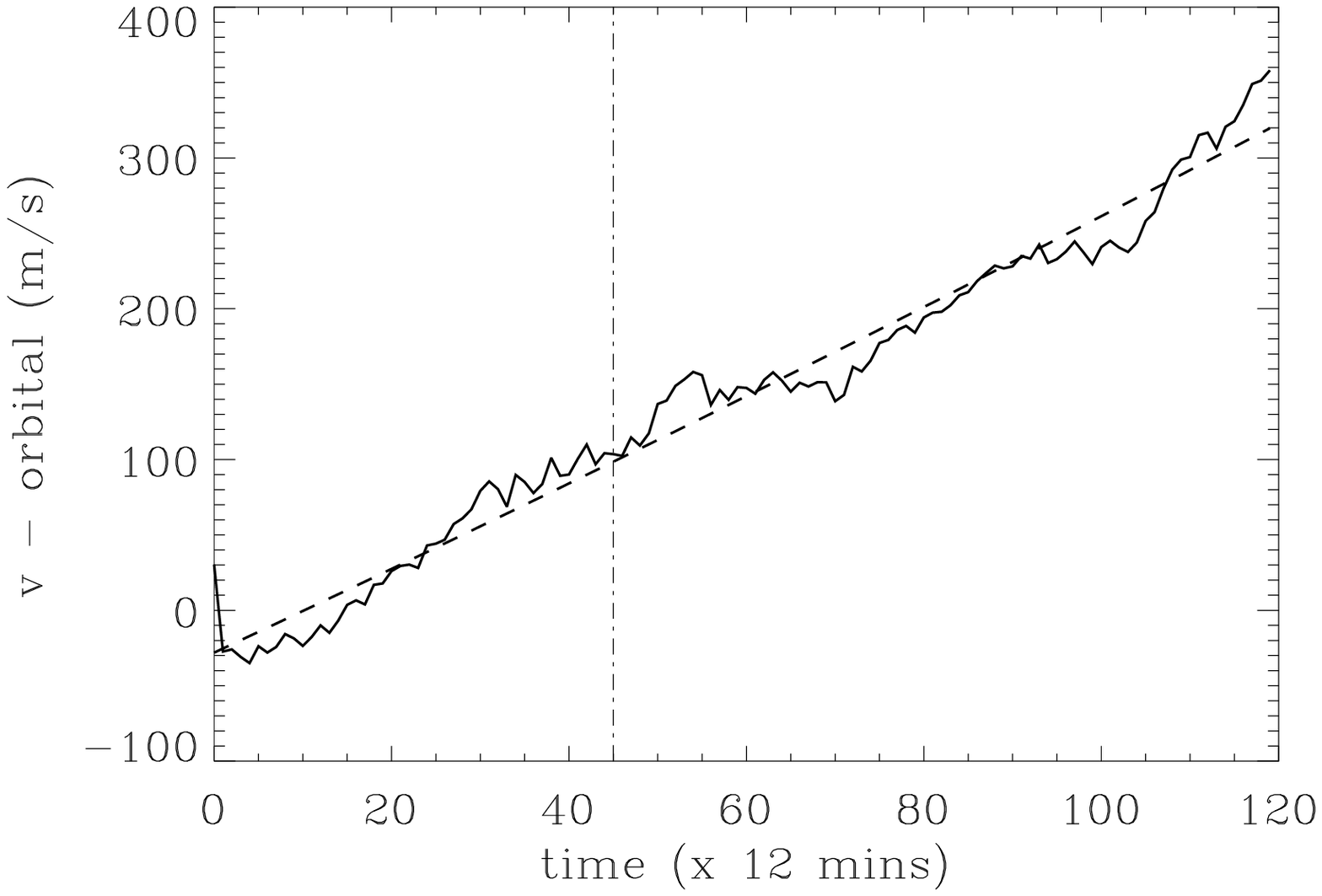}
\includegraphics[angle=0, scale=0.3]{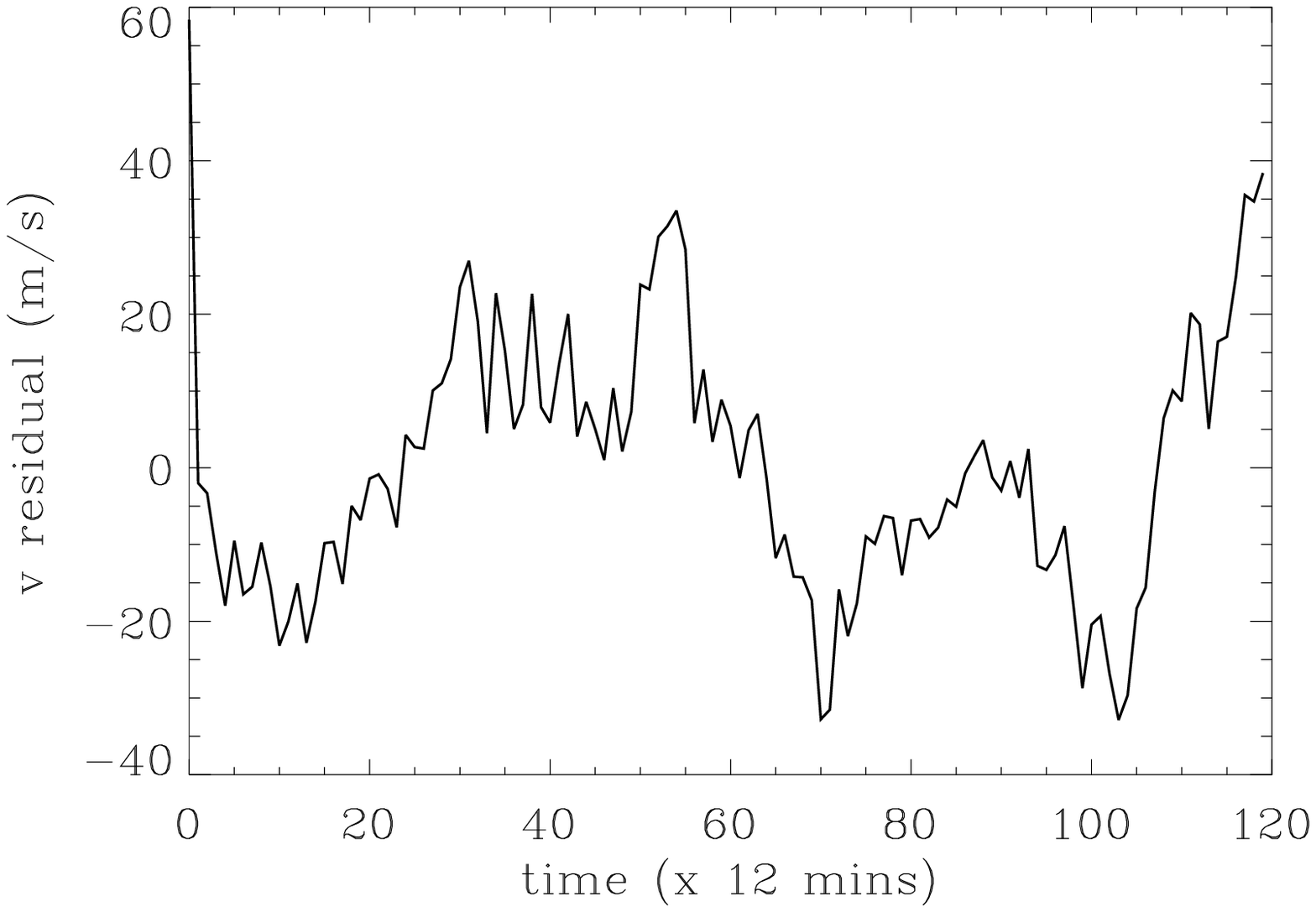}
\includegraphics[angle=0, scale=0.3]{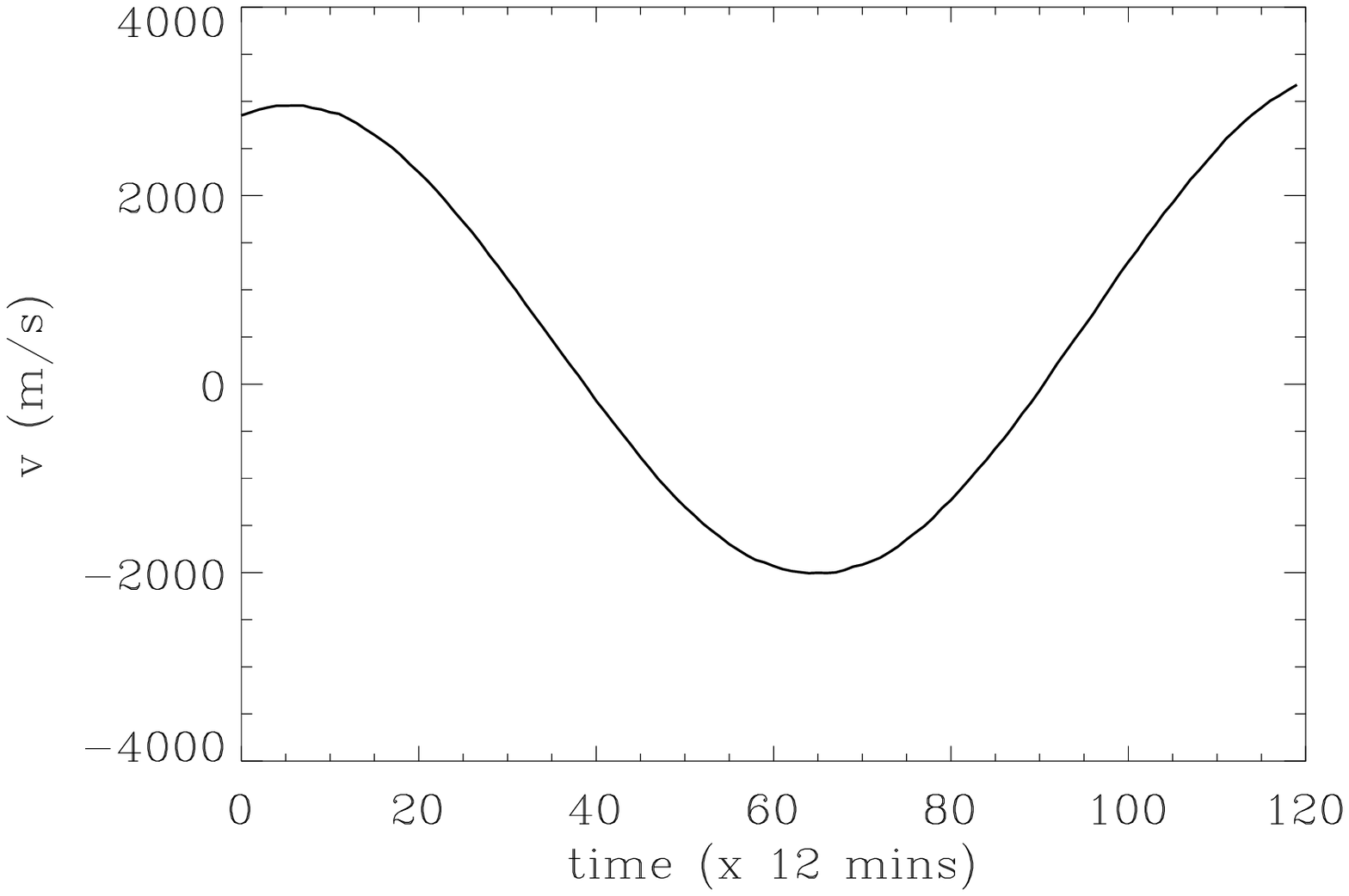}
\includegraphics[angle=0, scale=0.3]{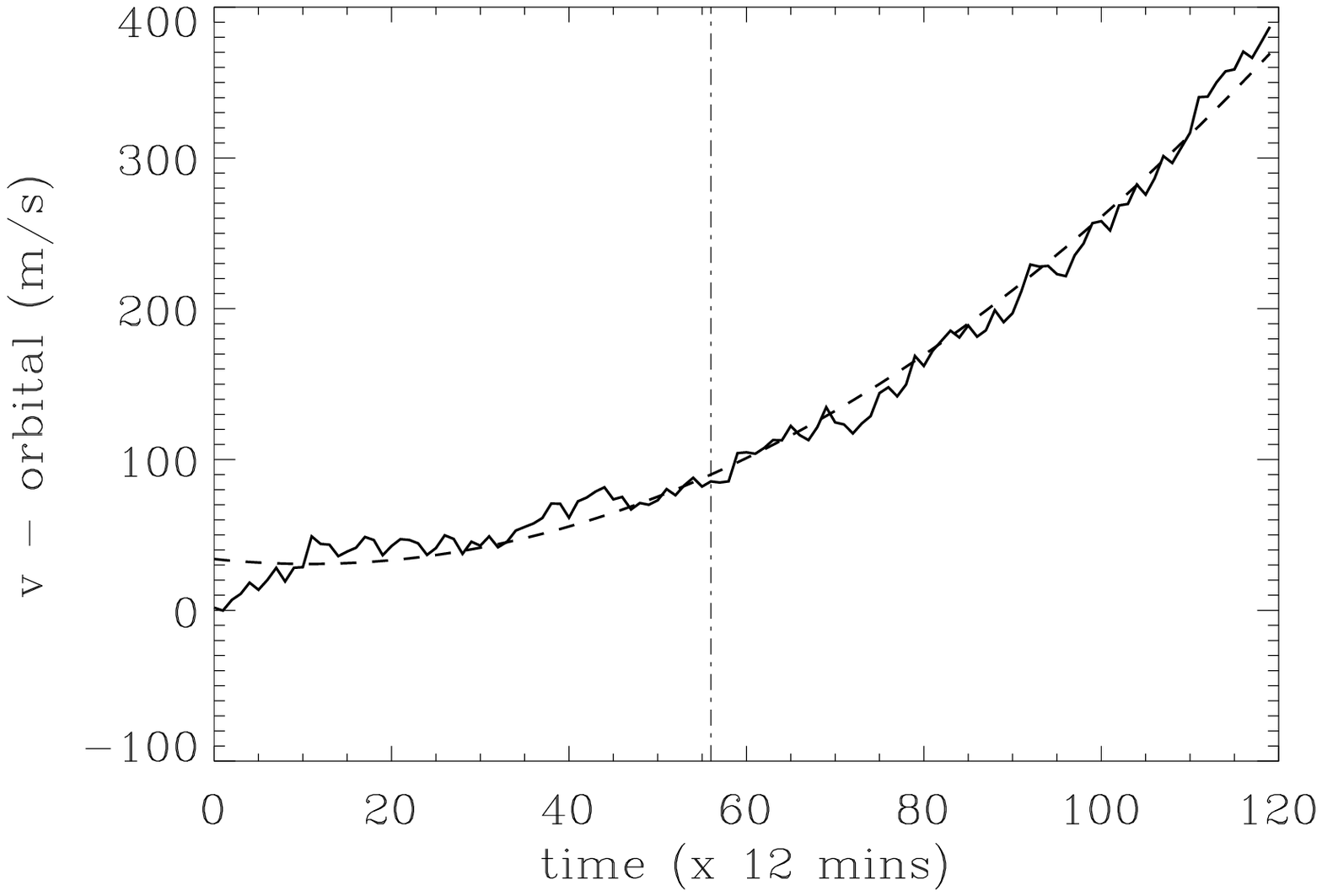}
\includegraphics[angle=0, scale=0.3]{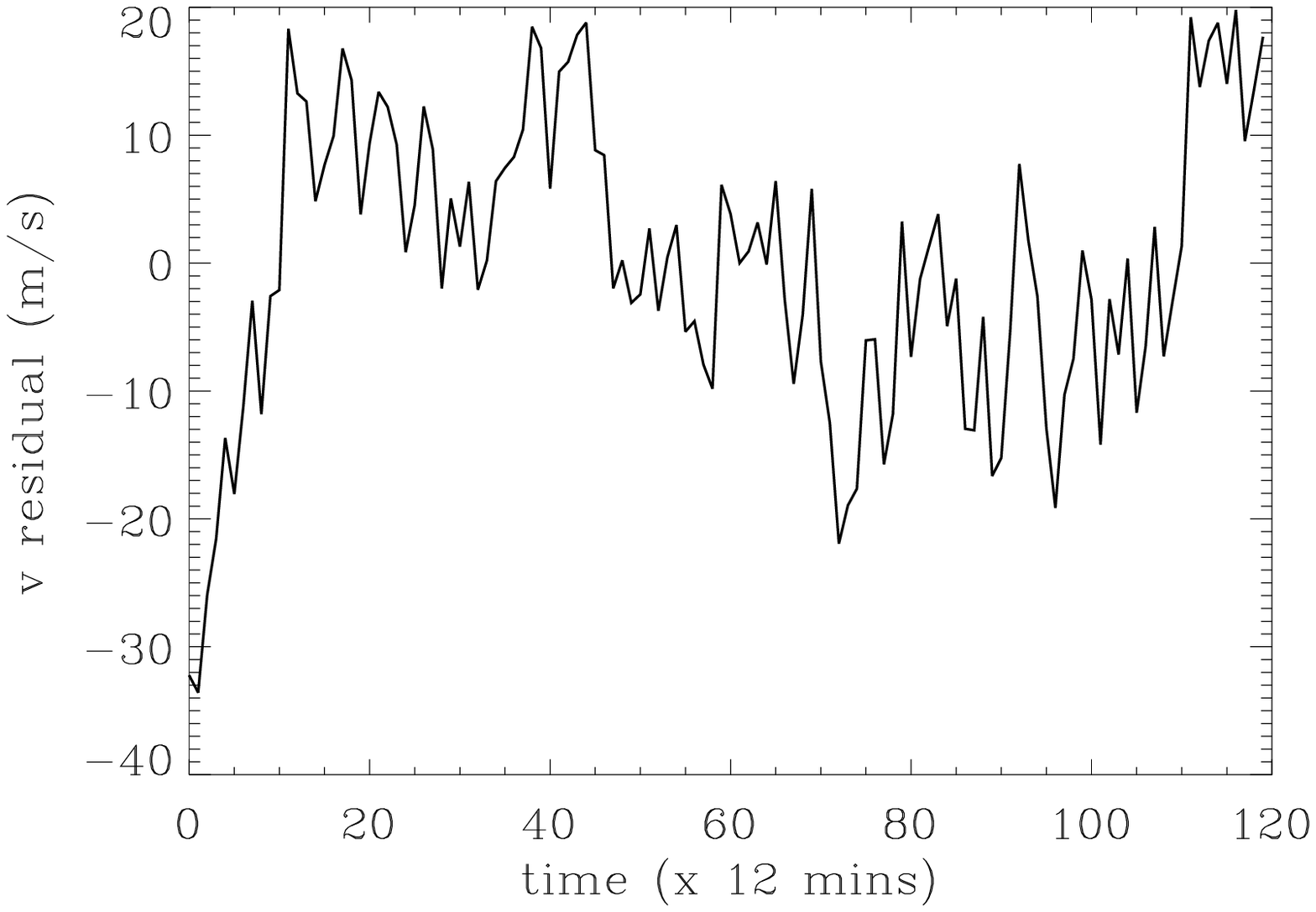}
\caption{Steps in the velocity calibration procedure for both datasets
  (AR 11105 at the top and AR 11211 at the bottom). The first column shows the mean velocity in the surrounding quiet Sun as a
  function of time. The middle column is what remains after removing
  the sinusoidal component due to the spacecraft velocity. A polynomial fit to
  this {\em solar} remnant (dashed line) is calculated, and the difference
  between the remnant and the fit is shown in the third column. The
  dashed-dotted line in the middle column marks the time at
  which the center of the AR crosses the central meridian.\label{fig:veloc_calib}}
\end{center}
\end{figure}

The kinematics of the observed features are important towards
understanding the 
emergence scenario of active regions. This requires an absolute calibration
of the Doppler velocities which, for an instrument like HMI,
that lacks an absolute wavelength reference such as a telluric line, is almost
impossible. 
The HMI filter profiles are centered at the Fe {\sc i} 6173 \AA\ wavelength
at rest, measured in vacuum and corrected for gravitational
redshift. This yields a reference wavelength of 6173.34 \AA\ \citep{norton2006}. However,
the instrument filter profiles are not tuned to follow the orbital velocity of the satellite, so
the first order component of the Doppler velocitiy inferred from the
spectral line inversions is a
sinusoid with a 24 hour period and an amplitude of approximately $\pm 3.5 {\rm kms}^{-1}$.

Our approach in this work is to calibrate the velocities in each AR
with respect to its non-magnetic surroundings. To this end, 
we take two strips of data, one North and one South of the AR, and
calculate the mean Doppler velocity in the combined areas as a function of time (first
column of fig. \ref{fig:veloc_calib}). Then, we subtract the effect
of the orbital motion of the SDO satellite. The difference (solid line in
middle column of fig. \ref{fig:veloc_calib}) comes mainly from solar
contributions, such as the center to limb variation of the
convective blueshift and the solar rotation as the AR travels from
East to West (although the former should be a
very small effect because the target ARs were never very far away from disk
center). There are other minor contributions
such a residual produced by the fringe pattern of the entrance window of
the telescope (Schlichenmaier, private communication) and
spatio-temporal variations due to the change in the position of the
HMI transmittance filter profiles with respect to the rest wavelength
of the spectral line during the satellite orbit. In order to remove these contributions we 
fit the velocity difference of the middle panels to a second order polynomial (dashed line in the
middle column of fig. \ref{fig:veloc_calib}). After subtracting the
fit, a residual no
larger than $\pm 50 {\rm ms^{-1}}$ is left (right column of fig. \ref{fig:veloc_calib}). 

We apply this calibration, obtained from the non-magnetic areas,
to the AR Doppler velocities.
In this sense, we are not carrying out an absolute velocity
calibration. Instead, we obtain a consistent measure of whether
material is flowing upwards or downwards inside the AR with respect to its quiet, uneventful surroundings.

\section{Evolution during the first stages of emergence}\label{sec:evolution}

\begin{figure}
\begin{center}
\includegraphics[angle=0, scale=0.4]{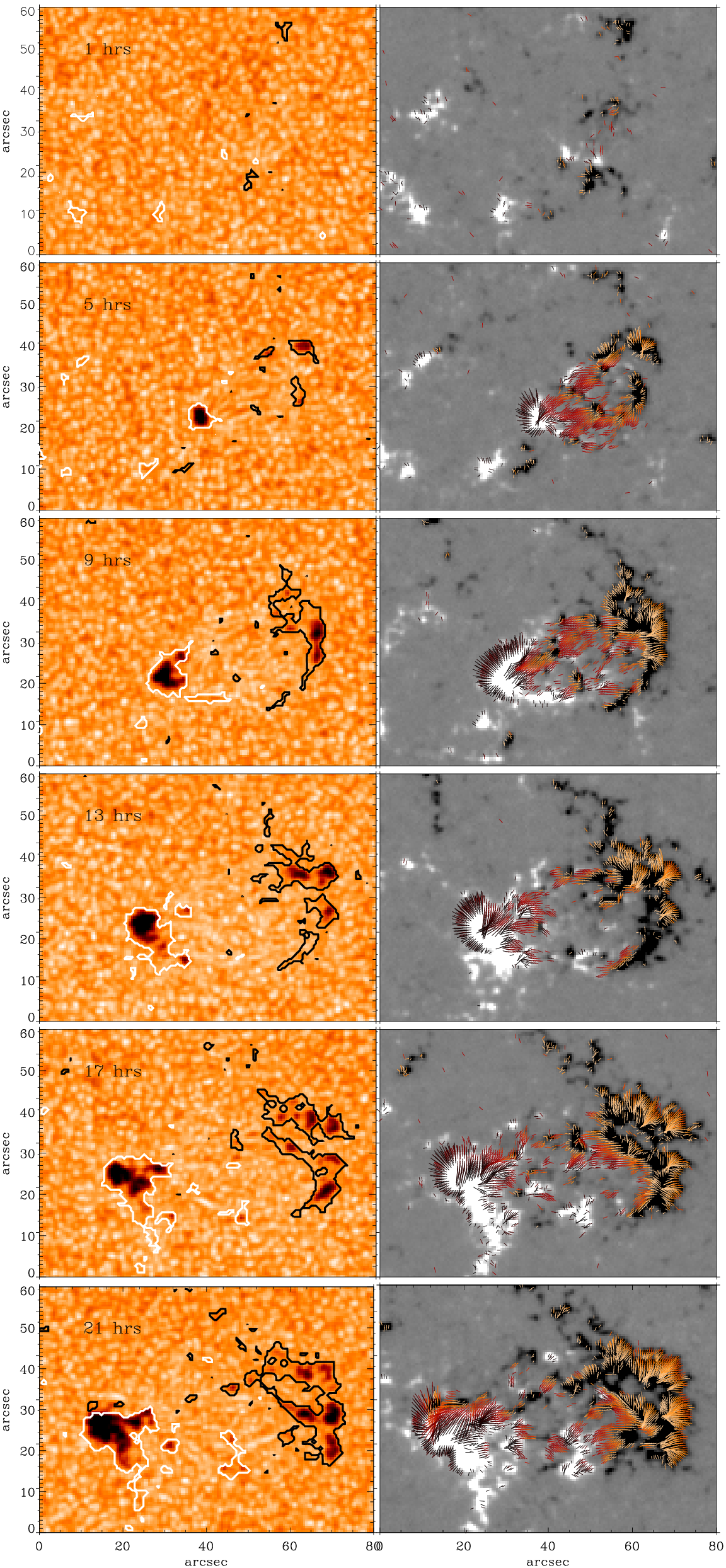}
\includegraphics[angle=0, scale=0.4]{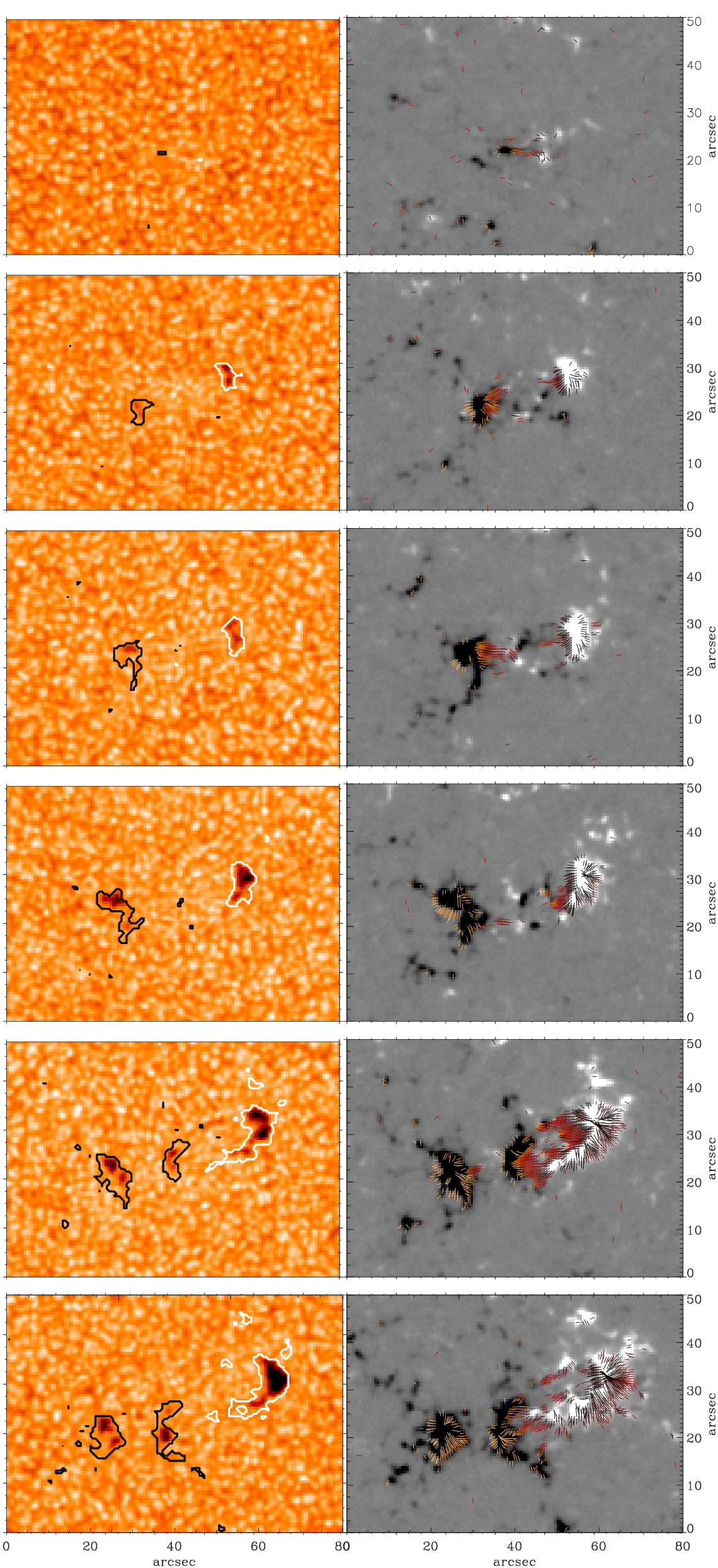}
\caption{Sequence of the emergence of AR 11105 on the left (first two columns) and AR
  11211 on the right. Time increases downwards, showing one image every
 4 hours. For each data-set, the first column represents the continuum
  intensity with the black and white contours showing $B{_{LOS}}$ at
  $\pm$150G. The second column shows $B_{LOS}$ as a gray-scale
  background saturated between -150 and 150 G (black/white represent
  negative/positive polarities). Superposed, the
  headless arrows represent the transverse component of the magnetic
  field, $B_{T}$, only for pixels in which $B_{T}$ is
  above the noise level and the inclination is in the $50^{\circ}-130^{\circ}$
  range. The black-red-white colour coding indicates values in this
  range, with red corresponding to purely transverse fields ($90^{\circ}$). \label{fig:evol}}
\end{center}
\end{figure}

Fig \ref{fig:evol} shows the sequence of the emergence for both
data-sets. Time increases downwards and only one snapshot every 4
hours 
is shown. The first two columns correspond to AR 11105 and the other
two to AR 11211. For each data-set, the first column represents the
continuum intensity in the background with white and black contours
showing the line-of-sight magnetic flux density, ${\rm B_{LOS}}$, at $\pm 150\,$G.
The second column shows ${\rm B_{LOS}}$ as the grayscale background
with superposed (headless) arrows representing the direction of the transverse
component of the vector magnetic field, ${\rm B_{T}}$. Arrows are only
plotted for pixels whose magnetic field is above the noise level and mostly horizontal with respect to the solar surface
 (i.e. with inclinations $\theta_B$, between $50-130^{\circ}$).

AR 11105 emerges in the Northern Hemisphere on September 1 2010. It grows fast, at large flux rate. Over the
first few days of its life it develops many pores that coalesce into
larger spots that eventually develop penumbrae. By the time 
it reaches the West limb, AR 11105 is composed of a large leading sunspot
followed by groups of pores and surrounded by strong plage areas.
AR 11211, on the other hand, appears in the Southern Hemisphere on May 8 2011. It only develops a few small pores that decay
within three days of emerging. After that, its signature disappears completely from the continuum images,
only to be seen in the magnetograms. Any magnetic remnant of this region fades away
before crossing the West limb.

In this paper we focus on the 24 hours following the first signatures
of the emergence. A rough common scenario applies to both ARs:
during the first minutes of the emergence process, the only
signatures of the AR are a small patch of horizontal fields connecting
a magnetic dipole. In the following hours, new bursts of transverse fields
bring more flux to the surface. As they rise, the anchoring points
of the newly emerged field lines become more and more vertical, and drift
and merge with pre-existing flux of the same polarity. The main footpoints
of the active regions begin to differentiate themselves from the
surrounding flux as they drift apart. After 2-3 hours, the first pores
start to appear in the continuum images. The AR carries on growing 
without interruption, bringing magnetic flux to the surface at a
steady rate.
One of the differenes between the two ARs is the development of
abnormal granulation in the area where the horizontal fields
come to the surface. AR 11105 (left columns of figure
\ref{fig:evol}) exhibits this characteristic whenever a new patch of
horizontal field appears in the vector-magnetogram - with the granulation
becoming brighter, with less contrast and elongated in the direction of the magnetic field. This
effect, albeit also present in AR 11211, is not as common or
as strong as in the first case (right columns of figure
\ref{fig:evol}), probably because of the weaker nature of the
magnetic fields that rise through the surface.

\noindent This paper will focus on the magnetic and kinematic
properties of naked AR emergence, and the relation between them during these first stages of the emergence.

\subsection{Flux history and other magnetic quantities}
\begin{figure}[!t]
\begin{center}
\includegraphics[angle=0, scale=0.45]{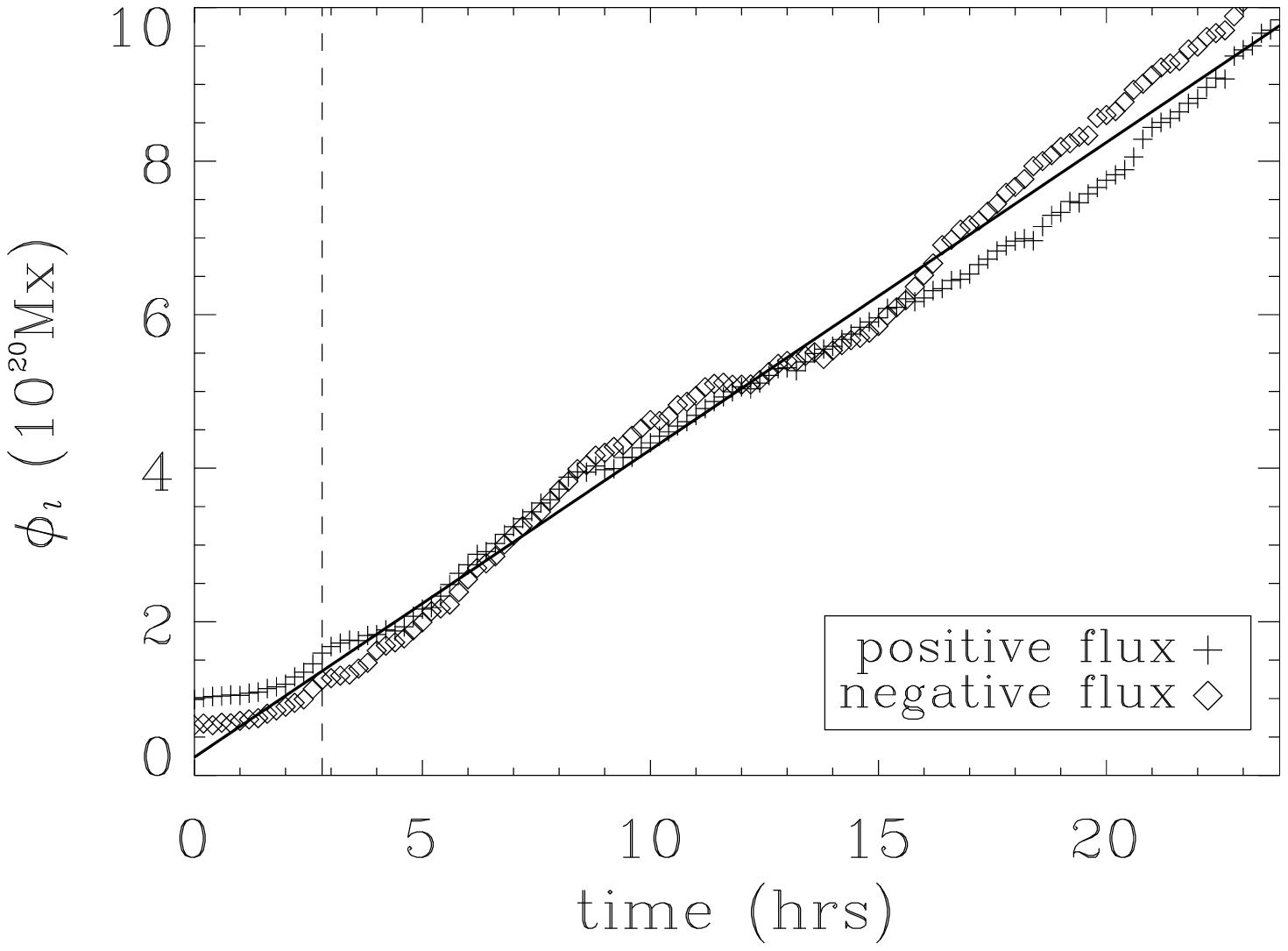}
\includegraphics[angle=0, scale=0.45]{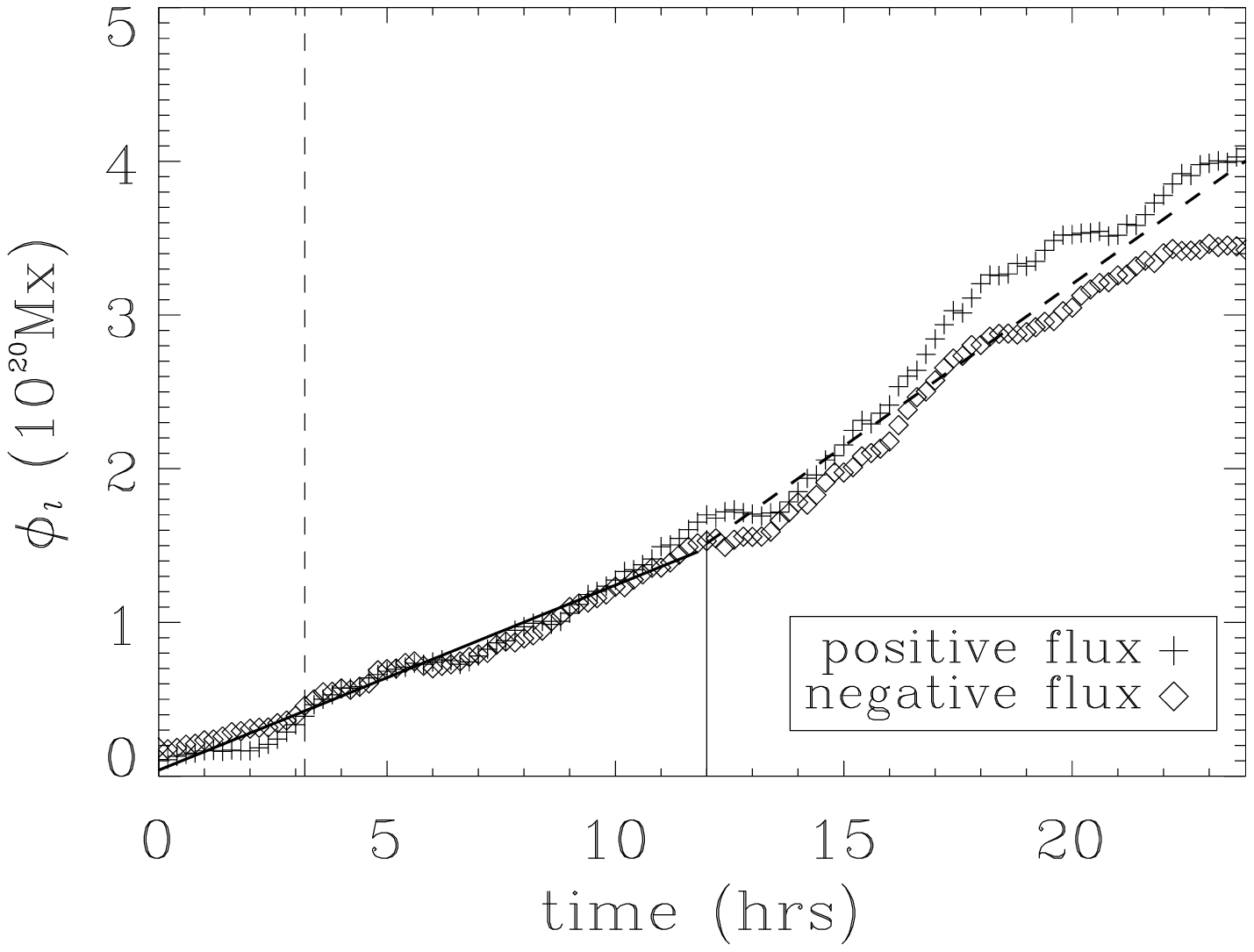}
\caption{Flux history for both active regions, NOAA AR 11105 on the
  left and AR 11211 on the right. The dashed vertical line shows the
  time at which the first pores appear in the continuum intensity. 
In the first case, the flux rate is $\sim 4\cdot10^{19}$ Mx/hr while in the second one, it varies from
  $1.2$ to $2 \cdot 10^{19}$ Mx/hr. Two different linear fits have been
  applied to the flux evolution of AR 11211, in the first and the
  second halves of the sequence. This somewhat arbitrary division is
  marked by the solid vertical line in the right panel of the figure.\label{fig:flux}}
\end{center}
\end{figure}

Figure \ref{fig:flux} shows the flux history for both regions during
the first 24 hours of existence. Diamonds represent the
negative component and plus signs the positive component of the
magnetic flux. In both cases, during the first 15 hours, the positive
and negative fluxes are very well
balanced. However, over time, the ARs occupy a larger area and a flux
imbalance becomes evident. In the case of AR 11211 (right panel),
towards the secont half of the sequence, the leading polarity (positive) becomes
more compact than the following one. Just because the following
polarity is more spread out, some of its
signal might lie below the detection threshold of the HMI instrument,
leading to a slight imbalance between the measured fluxes.
This is not the case for AR 11105 (left), whose positive flux hits the
edges of the FOV around hour 16, leading to a loss of flux through the
boundaries of the box, and hence to an imbalance in the flux curve.

 A linear fit to the flux curves shows that, during the first 12 hours
of emergence, the rate of flux brought to the surface by the large
active region (AR 11105 at $4\cdot 10^{19}$Mx/hr) triples that of the
smaller one (AR 11211). However, the latter increases
its flux output to $\sim 2\cdot 10^{19}$Mx/hr during the second half
of the day. The solid vertical line at minute 60 marks, approximately,
the time when the flux output changes.
The vertical dahsed lines in figure \ref{fig:flux} show the
approximate time at which the first pores are seen in the continuum intensity.

\begin{figure}[!t]
\begin{center}
\includegraphics[angle=0, scale=0.45]{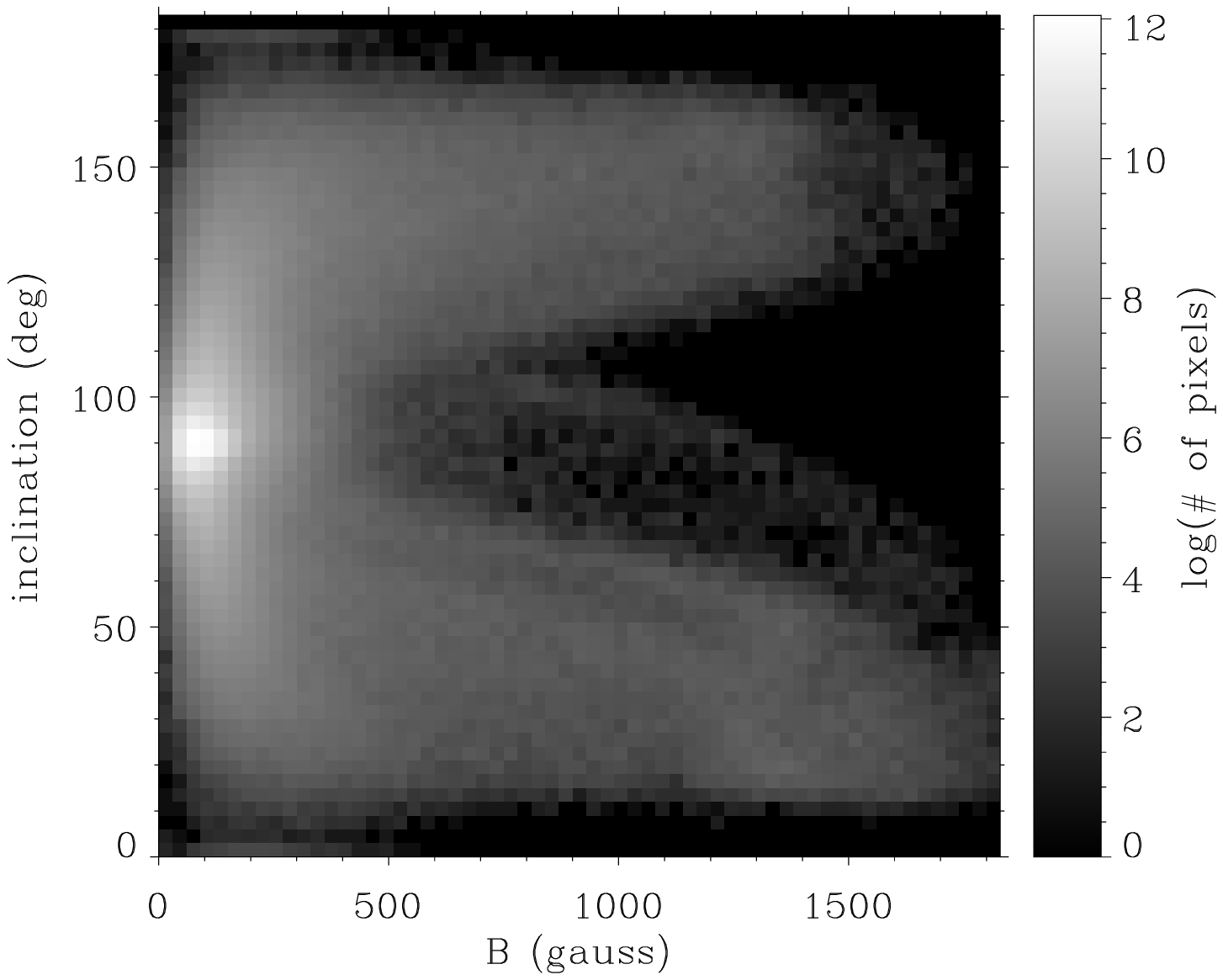}
\includegraphics[angle=0, scale=0.45]{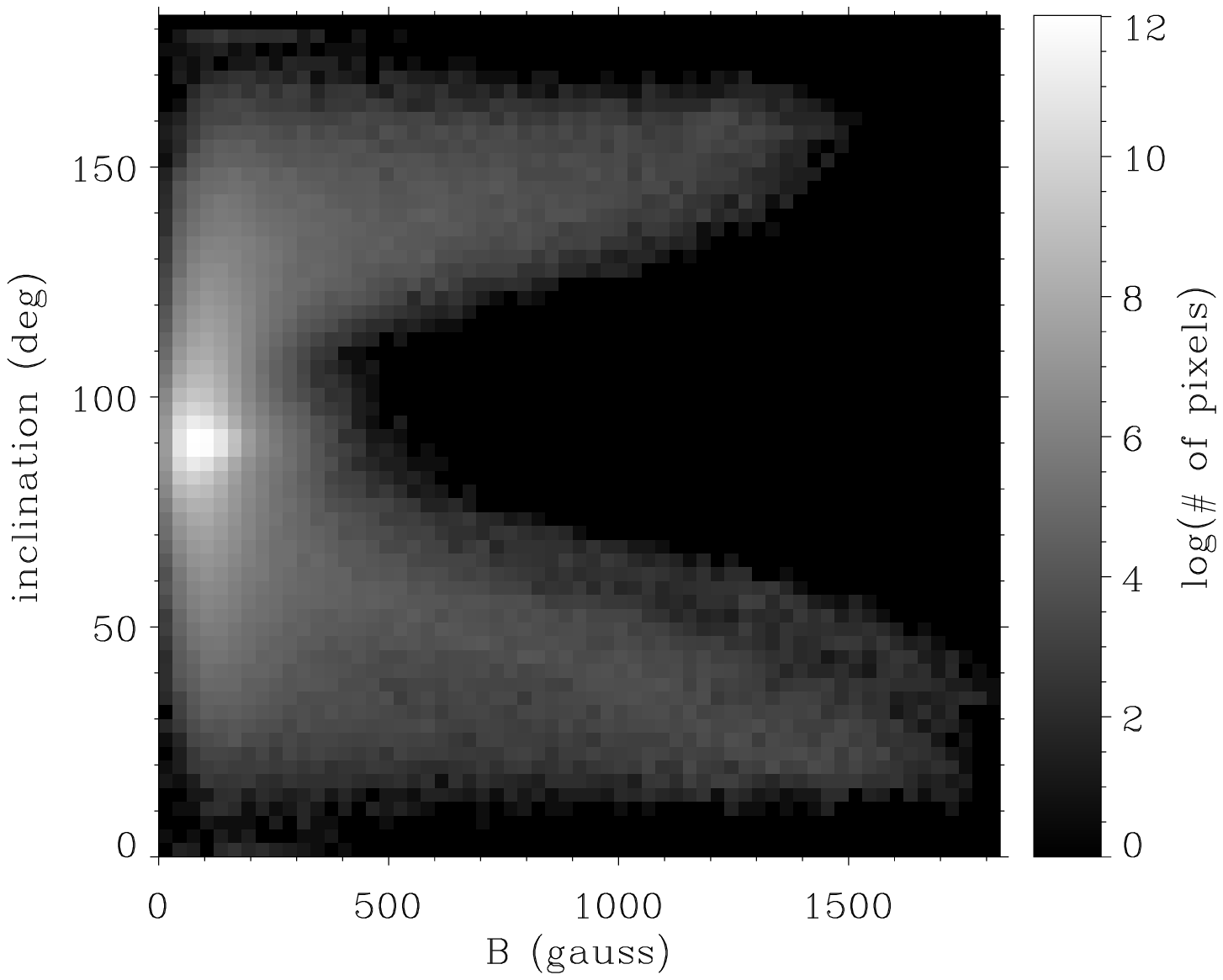}
\caption{Density scatter plots of the inverted magnetic field
  inclination vs. the field strength for both ARs during the first 24 hours 
of emergence. In both cases, the stronger fields are the most vertical
  ones. The strength of the transverse fields lies in between
  150-1000\,G for AR 11105 (left) and 150-600\,G for AR 11211
  (right). Note that everything below 150\,G is not reliable.\label{fig:strength}}
\end{center}
\end{figure}

Figure \ref{fig:strength} shows a scatter plot of the magnetic field
inclination (with respect to the LOS) vs. the field strength for the
combined first 24 hours of emergence. AR 11105 is on the left and AR
11211 on the right. Despite being represented in these figures,
transverse fields ($\sim 90^{\circ}$) with strengths below 150\,G are
the result of the inversion of polarization signals that are pure noise. 
In fact, the strongest signal in the scatter plots
correspond to this ``horizontal field effect'' of the spectral line
inversion, and it has no physical meaning or significance.
For both ARs, the
strongest fields tend to be mostly vertical. Transverse fields (around
90 degrees of inclination) are comparatively much weaker, ranging between
150-900\,G for AR 11105 and 150-400\,G for AR 11211.

AR 11105 grows faster and at a higher flux rate than AR
11211. The larger flux rate is not only a consequence of the
larger area, but also of the 
stronger transverse fields that it brings to the surface. Although the 
spectral line inversion that we use does not account for a magnetic
filling factor and hence does not deliver intrinsic field strengths, \citet{kubo2003}
stated that the filling factors in emergence zones are always larger than
80\%, which would mean that the strengths that we report here should be close to the intrinsic values.
AR 11105 also harbours stronger
vertical fields at its footpoints than AR 11211.

\subsection{Moving dipolar features}

\begin{figure}
\begin{center}
\includegraphics[angle=0, scale=0.45]{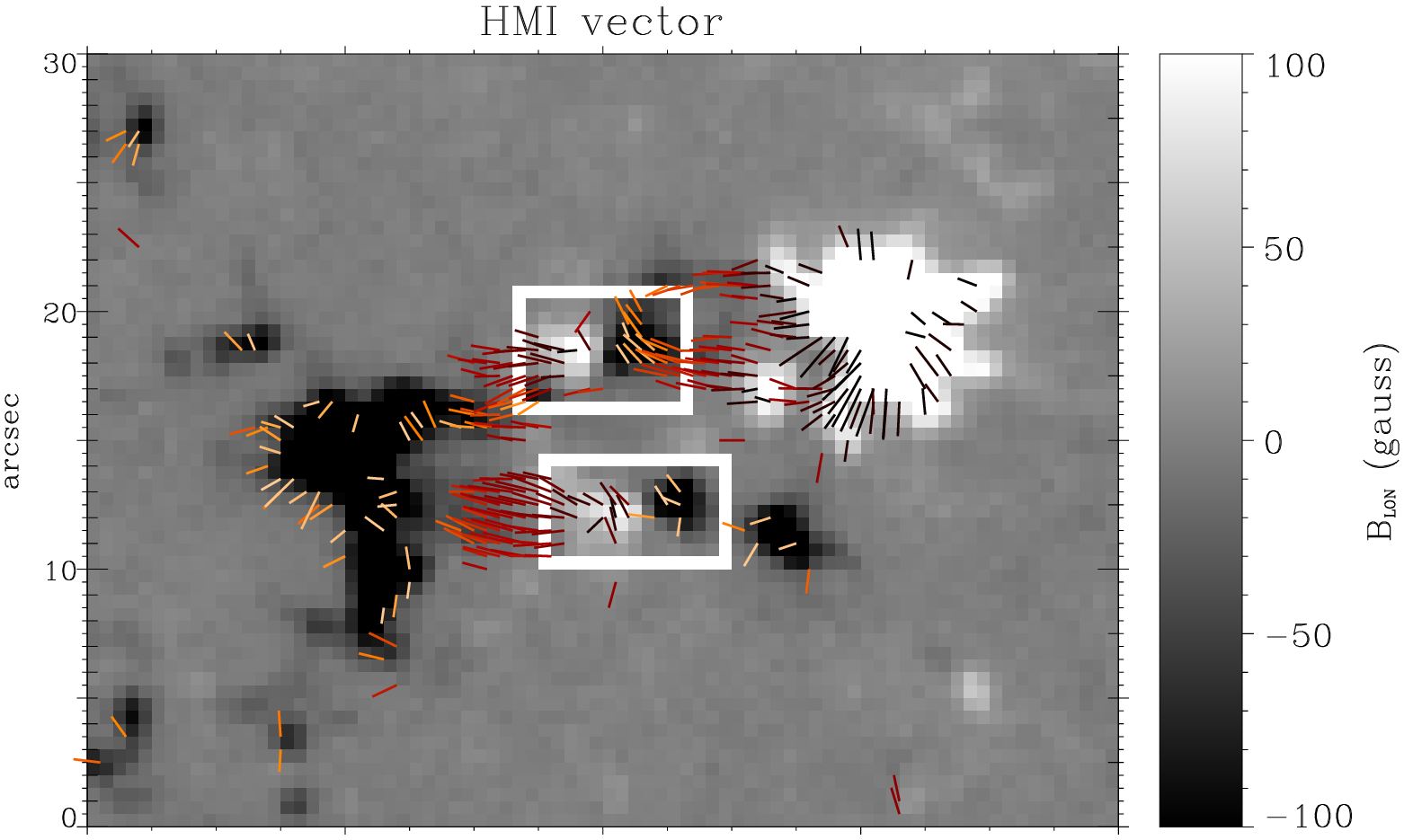}
\includegraphics[angle=0, scale=0.45]{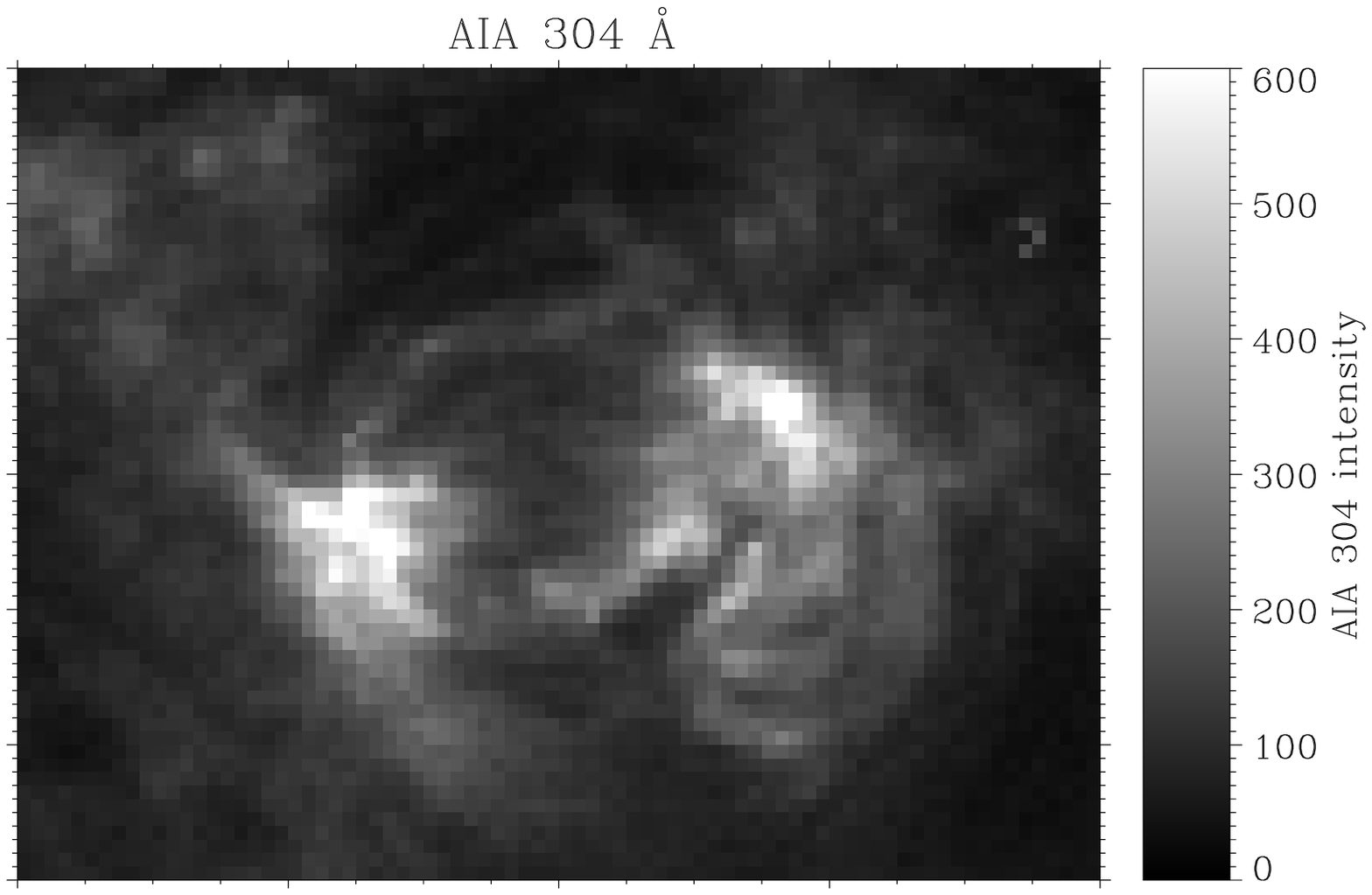} \\
\includegraphics[angle=0, scale=0.45]{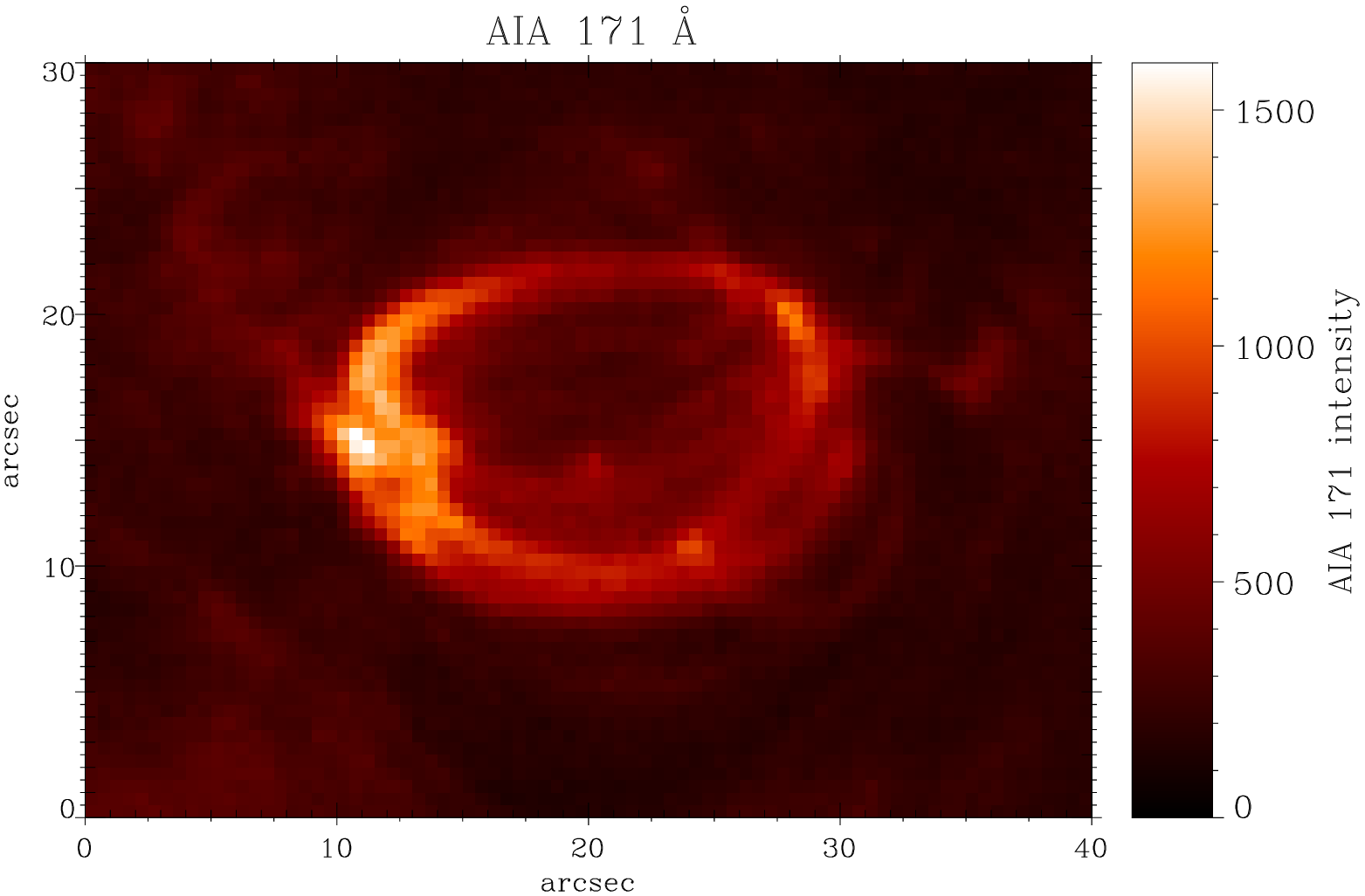}
\includegraphics[angle=0, scale=0.45]{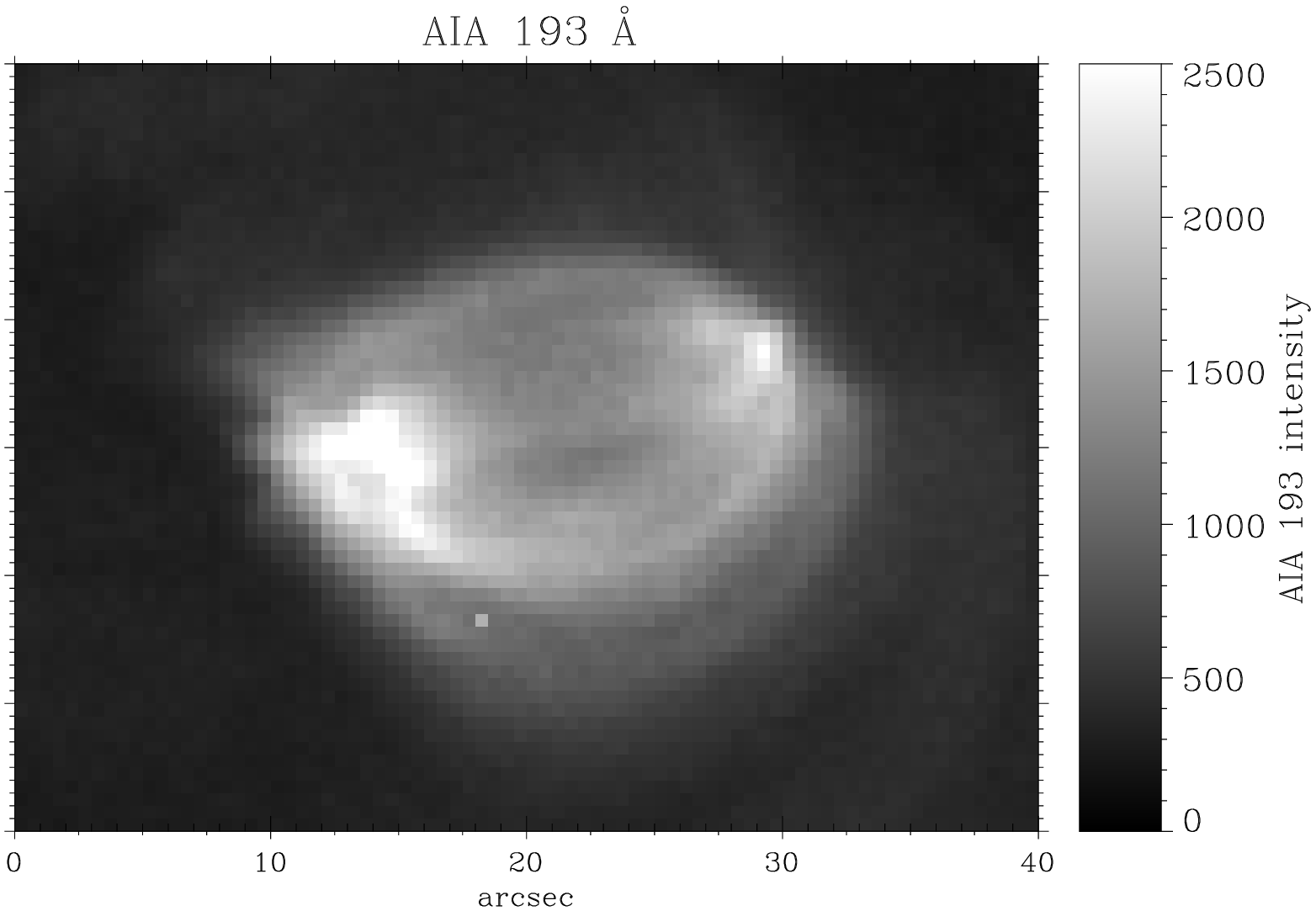} 
\caption{Top left: Detail of the magnetic field configuration around two MDFs
  (boxed in white) for AR 11211. As usual, the grayscale background
  represents the longitudinal magnetic flux density (positive/negative
  in white/black) while the red-ish
  headless arrows show the transverse component of the magnetic field
  for inclinations between $50^{\circ}$ and $130^{\circ}$ with respect to the
  LOS. The polarities of both MDFs are connected to the main
  footpoints of the AR by horizontal fields, rendering a picture of
  field lines that serpentine in and out of the solar surface. 
Top right: Quasi-simultaneous AIA 304 \AA\ image that shows the loop structure in
the chromosphere. Bottom: Quasi-simultaneous AIA 171 \AA\ (left) and
AIA 193 \AA\ (right) channels
showing the structure in
the corona.\label{fig:mdf_detail}}
\end{center}
\end{figure}
During the first stages of the emergence process, horizontal magnetic
fields connect the footpoints
of the AR, which grow fast, drift apart and differentiate themselves as
the main dipole where the AR is anchored to the Photosphere. 
As soon as these footpoints are far enough (5-10''), the flux
emergence process starts happening within smaller patches of
horizontal fields that do not span the whole distance between the main
footpoints.
The top left panel of figure \ref{fig:mdf_detail} shows one snapshot in the emergence of AR
11211. The main footpoints of the AR can be easily seen in the
background magnetogram. In
between the main polarities, two small dipoles
\citep[labelled {\em moving dipolar features}, MDFs, by][]{2002bernasconi} of opposite polarity to
that of the AR, are highlighted by the white boxes. Each of the poles
of these MDFs, is connected to one of the main footpoints of the AR by transverse fields (red
headless arrows). The color coding of the arrows is such that red
indicates a purely transverse field ($\theta=90^{\circ}$), while darker or lighter
represent a range of inclinations, between $50^{\circ} \ {\rm and} \ 130^{\circ}$, white
being a magnetic field mostly pointing into the solar surface and black
pointing outwards (note that this black/white convention is opposite to
that of the magnetogram, for clarity purposes).
It has to be born in mind that these data have not 
been disambiguated. However, due to the proximity to disk center,
the magnetic field disambiguation would only have a minor effect on the
inclination angle and the projected transverse
component of $\vec B$, rendering a very similar picture to that of
Fig. \ref{fig:mdf_detail}. The relatively simple configuration of the AR
and the obvious connectivity between  each pair of magnetic poles
calls for a scenario where the field lines that emerge from the main
positive polarity, dip into the photosphere at the locations of the
MDFs just to re-emerge again and arch their way over to the negative
polarity. MDFs are just a more vertical counterpart of the horizontal field
lines represented by the arrows.

\begin{figure}
\begin{center}
\includegraphics[angle=0, scale=0.3]{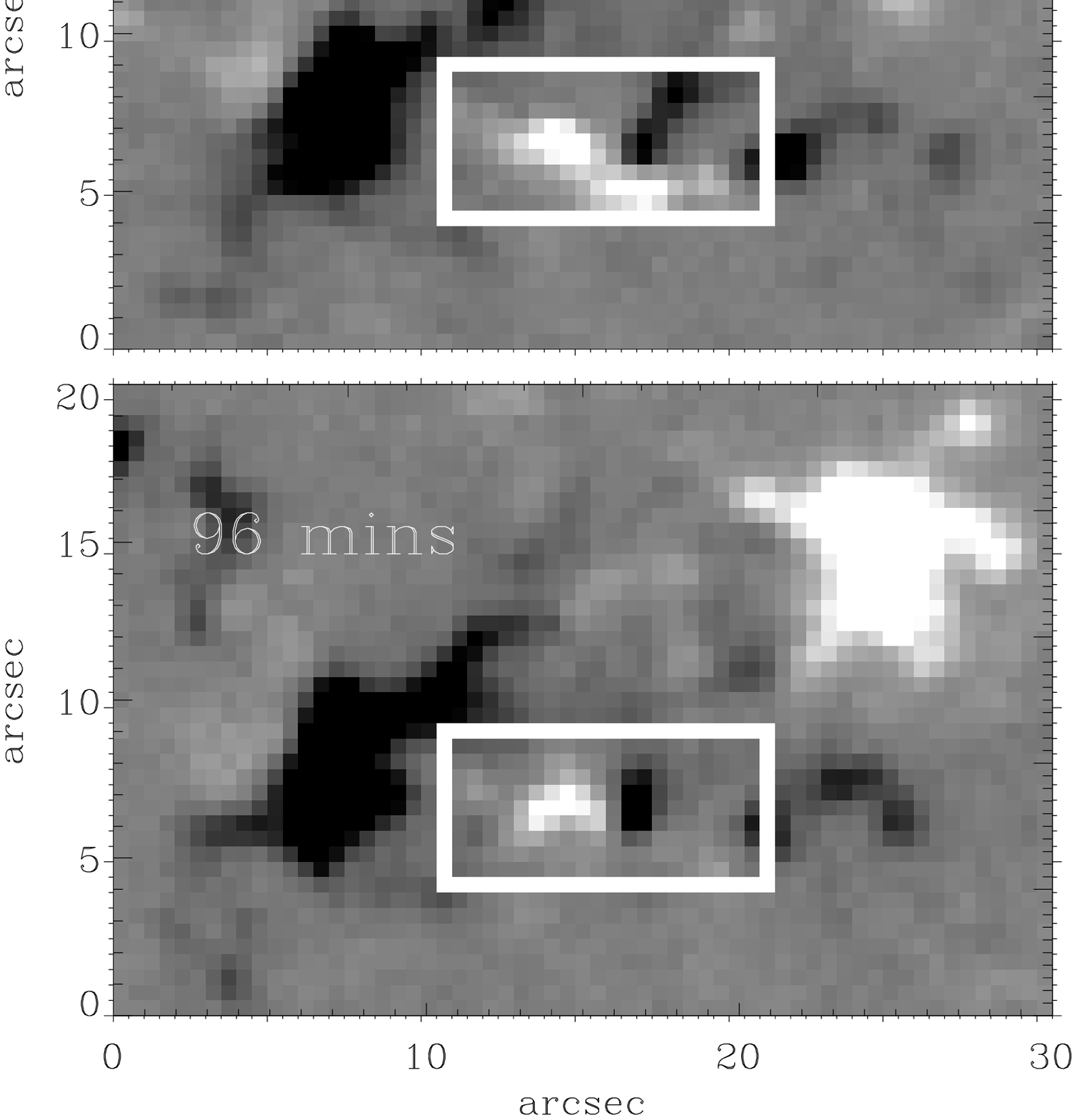}
\includegraphics[angle=0, scale=0.3]{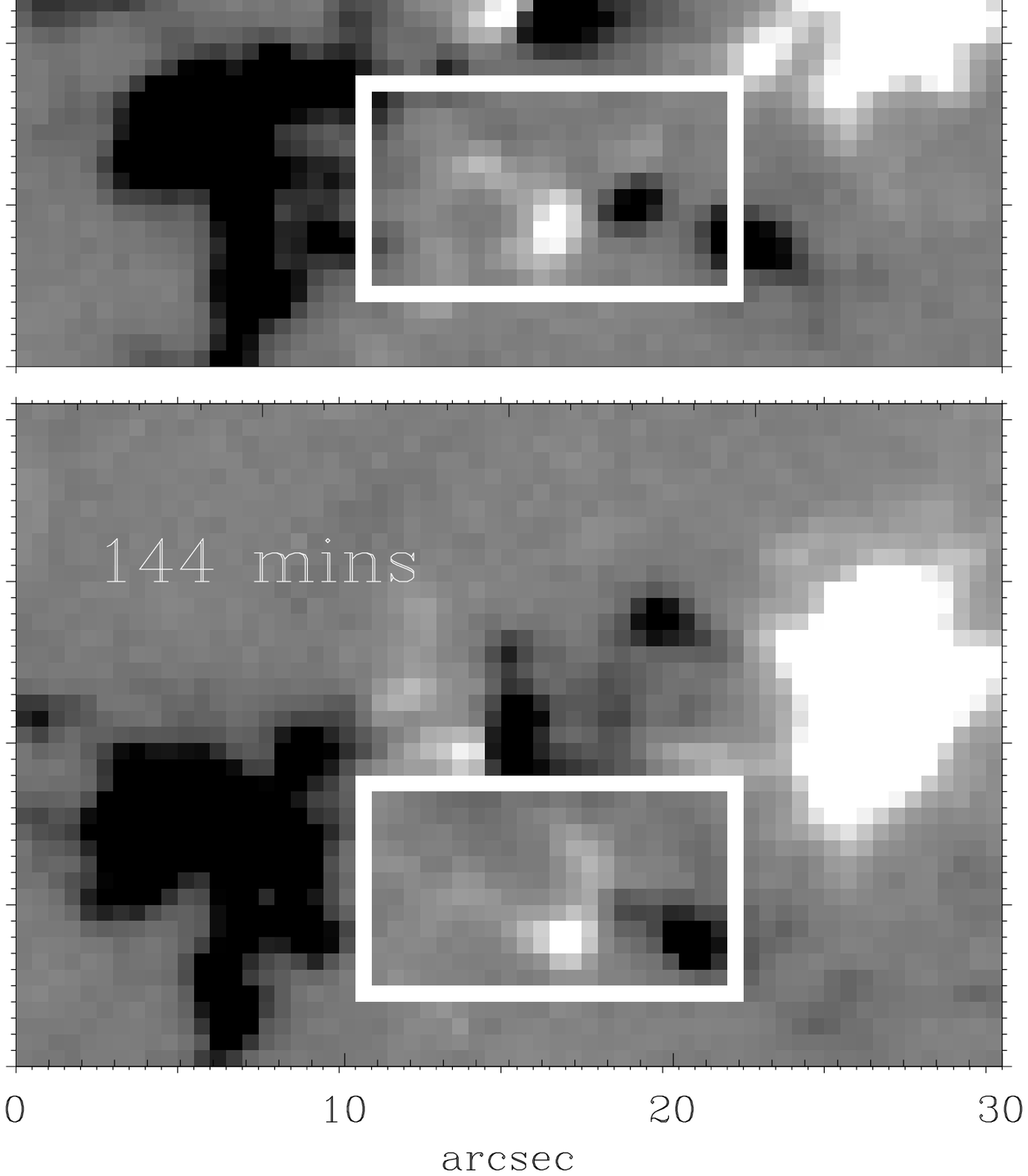}
\includegraphics[angle=0, scale=0.3]{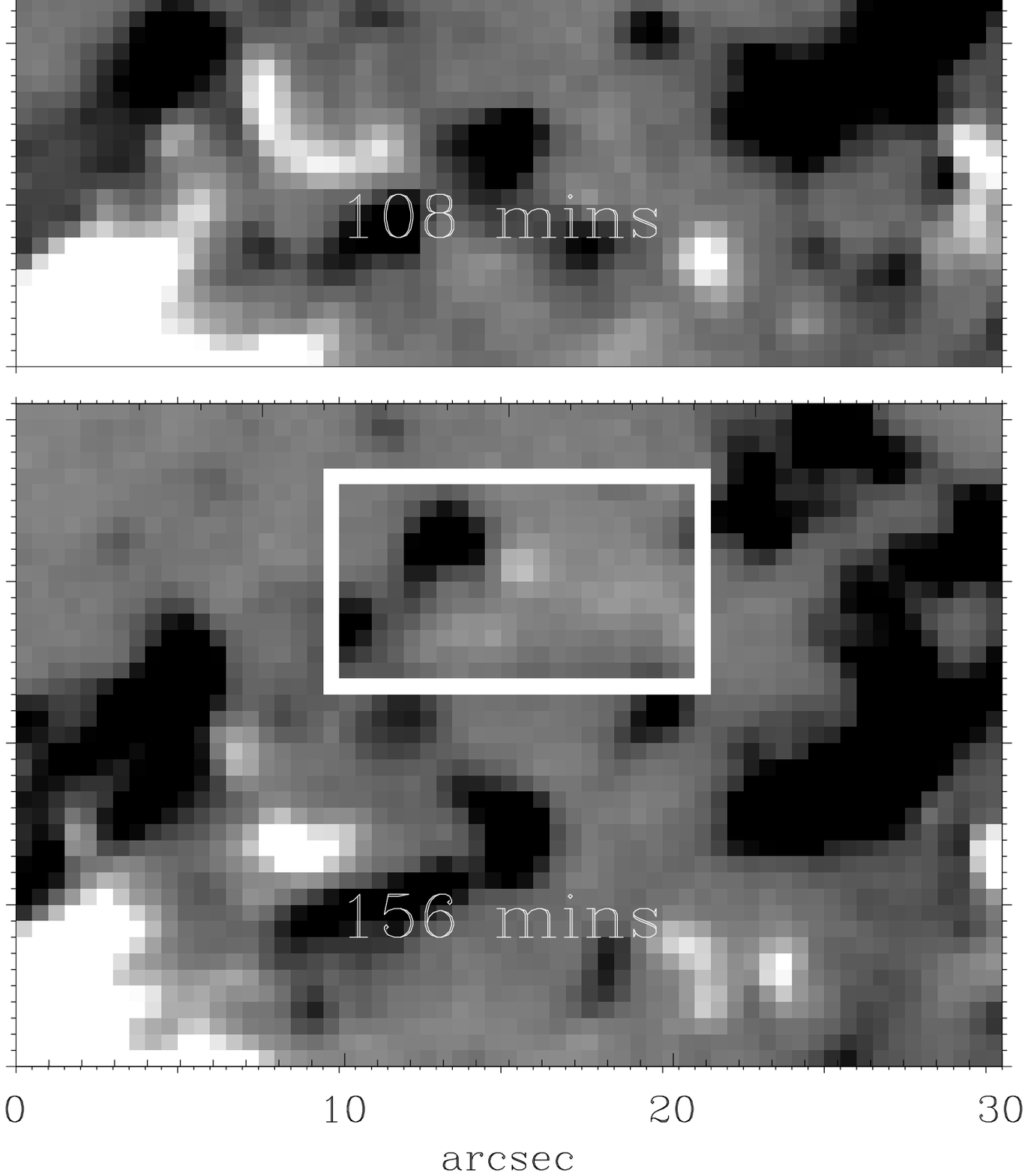}
\caption{Temporal evolution of three MDFs. The grayscale background
  in all panels shows the LOS magnetic flux density saturated at $\pm 100$\,G. 
The three columns represent
 three different events (the left and middle colums correspond to AR 11211
  and the right column to AR 11105). Time increases downwards, and the
  starting time is approximately 3 hrs, 6 hrs and 20 hrs (from left to
  right) with respect to the first signature of the emergence of the
  ARs. 
The white squares
  highlight isolated MDFs, whose footpoints come closer together in the
  time span of these sequences. For the sake of showing the final
  moments of the MDF (when one polarity has been mostly cancelled out
  by the other), the snapshots are not evenly spread out in time. \label{fig:mdfs}}
\end{center}
\end{figure}

 Fig. \ref{fig:mdfs} shows the temporal evolution of three individual
 MDFs, highlighted in the white boxes. The grayscale background shows
 the LOS magnetic flux density saturated at $\pm 100$\,G. Time
 increases downwards but
the snapshots are not necessarily evenly spread out in time.
As captured in this figure, the poles of the
MDFs often approach each other
until they merge, partially cancelling one another and sometimes completely
disappearing (the partial cancellation is obvious in the last panels
of the middle and the right columns of this figure). This behaviour is in contrast to what
\citet{2002bernasconi} see in their observations, namely that MDFs tend to
drift towards one of the main footpoints of the AR and then disappear.

\noindent The question of whether a reconnection event takes place
below or above the surface,
in between the footpoints of the MDF is up to debate. The HMI
measurements are blind to the layers
below or above the region of sensitivity of its spectral line,
hence any conjecture of what happens underneath the surface or in the high
the photosphere goes beyond the
capability of these observations and the scope of this paper.
If the reconnection happened above the photosphere, one would expect to see
downflows associated to the cancelling site, where the submerging part
of the magnetic loop crosses the surface. If, on the other hand, the reconnection took place
below the surface, ascending motions should be present where the
arcade finally manages to emerge. The HMI observations at a 12 minute
cadence do not provide evidence either way. The spatial and temporal
resolution and the uncertainties in the velocity measurements do not
offer an answer to this question. 

 In any
case, \citet{cheung2010} suggest this reconnection as a mechanism to
get rid of the mass carried by the rising field
lines and \citet{watanabe} even use it to explain the triggering of
Ellerman Bombs in the chromosphere.

MDFs are just the signature of emerging
field lines that are trapped in the photosphere due to entrained mass
that acts as an anchoring weight. Once an MDF cancels
out (supposedly after a reconnection of the dipped loop), the magnetic arcade is no longer pinned down to the photosphere
at these points and the loop will be free to rise into the
corona (while its counterpart submerges into the solar interior). 
The top right panel of figure \ref{fig:mdf_detail} shows a
near-simultaneous image of the chromosphere taken by the Atmospheric
Imaging Assembly \citep[AIA,][]{aia} in the 304 \AA\ wavelength. The bottom panels
of the same figure show, on the other hand, quasi-simultaneous images of the
corona in the AIA 171 and 193 \AA\ channels. While the chromospheric
image (at 304 \AA) presents
loops that are structured a lot like the photospheric field (the lower
left side of the image shows shorter loops connecting the lower MDF to
the leading footpoint of the AR), the
coronal structures span the entire distance between the main
footpoints, and are almost oblivious to what happens at the smaller spatial
scales seen at photospheric levels. 
That said, time sequences of AIA 171 and 193 \AA, show {\em transient} brightenings and
small loop-like structures associated to the locations of the MDFs
(the simultaneity with the photospheric phenomena cannot be pinned
down since the HMI vector data have a much lower cadence). However,
most of the time, the overall structure of the coronal images is that of
large-scale loops connecting the main footpoints of the AR.
These pieces of evidence support a
picture in which the emerging horizontal fields pinned down to the
photosphere at the MDF locations cannot rise much beyond the chromosphere
until the MDF reconnects and the magnetic arcade is released from its photospheric trap.

\subsection{Doppler velocities}
\begin{figure}[t!]
\begin{center}
\includegraphics[angle=0, scale=0.3]{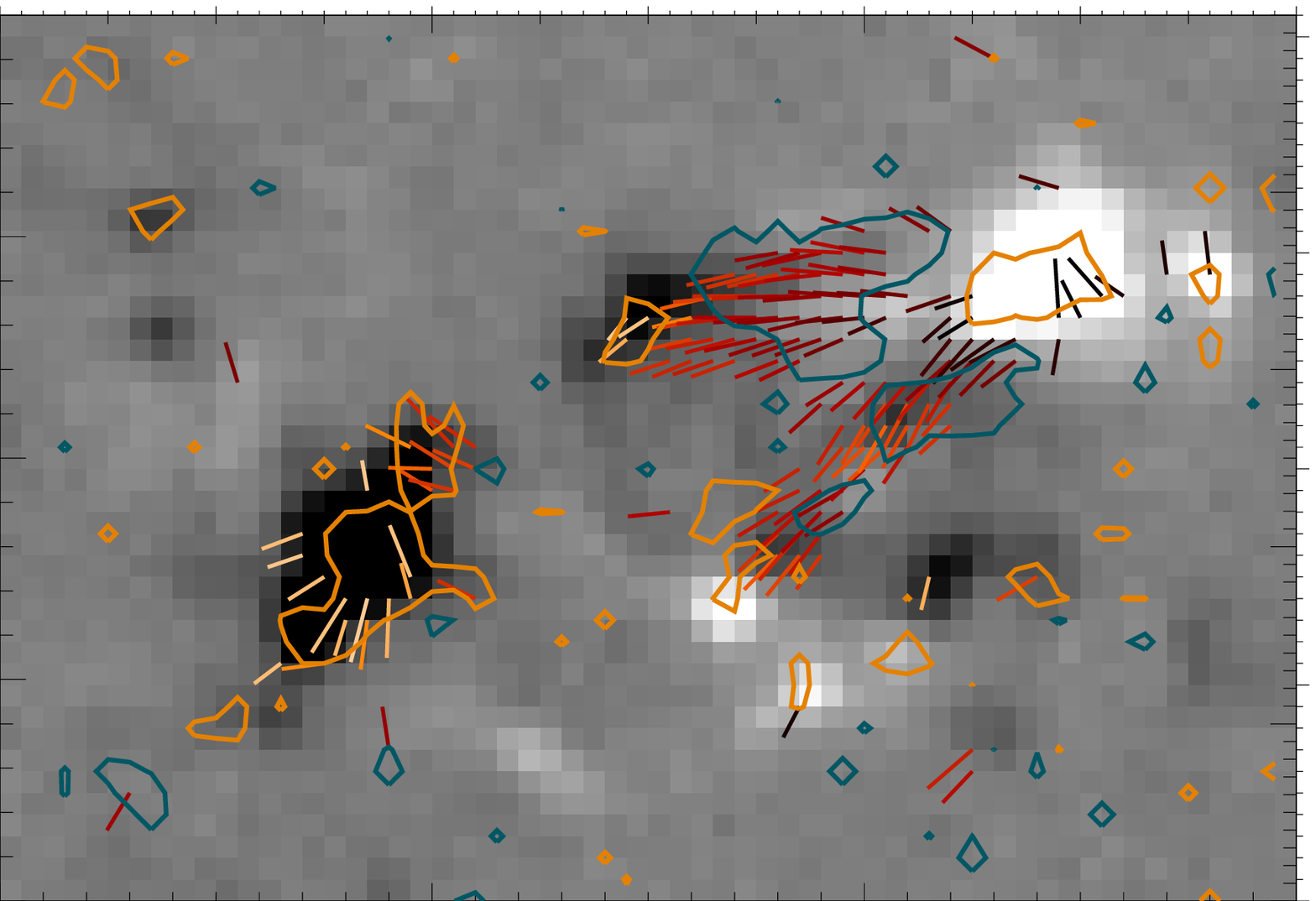}
\includegraphics[angle=0, scale=0.3]{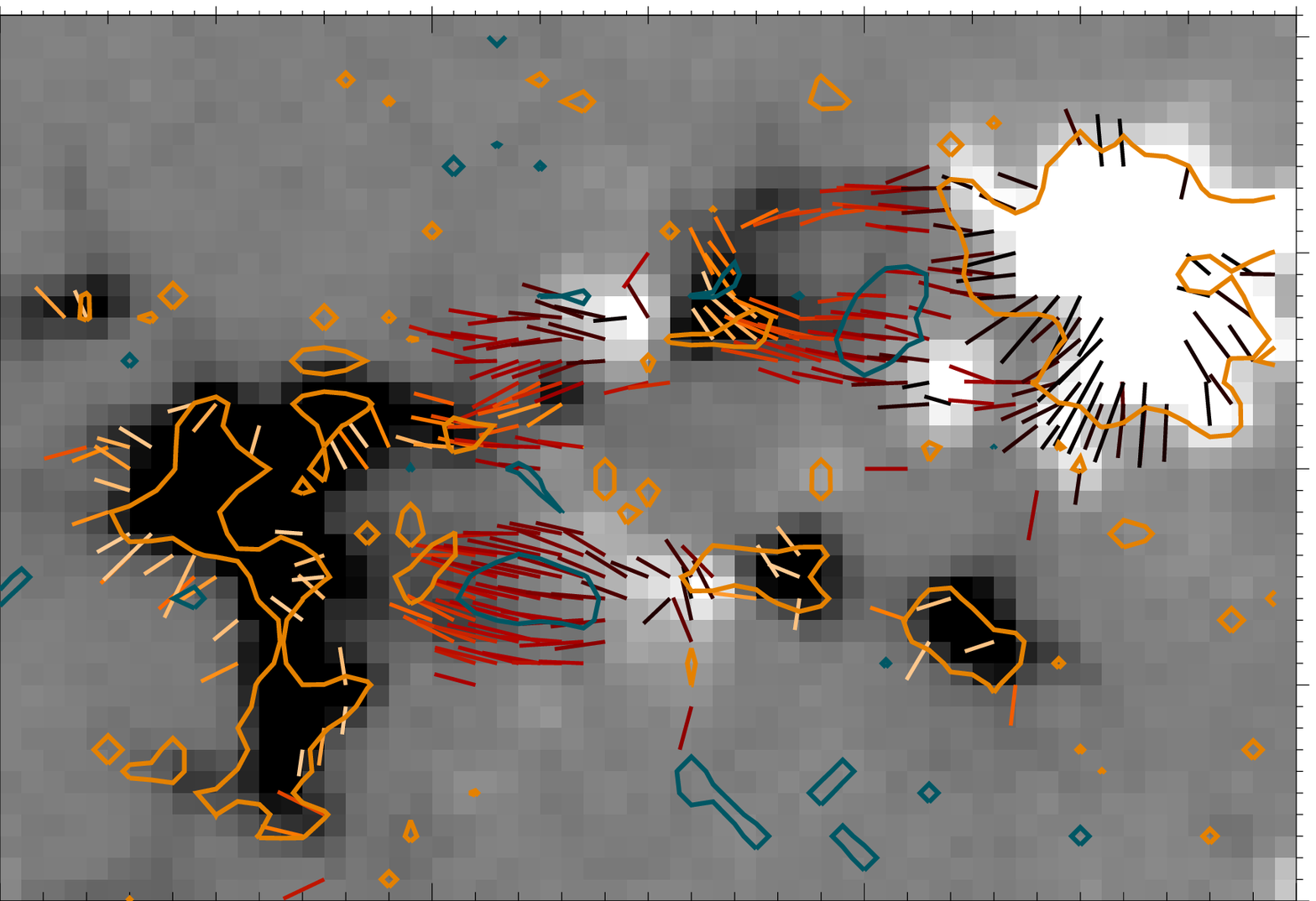}
\includegraphics[angle=0, scale=0.3]{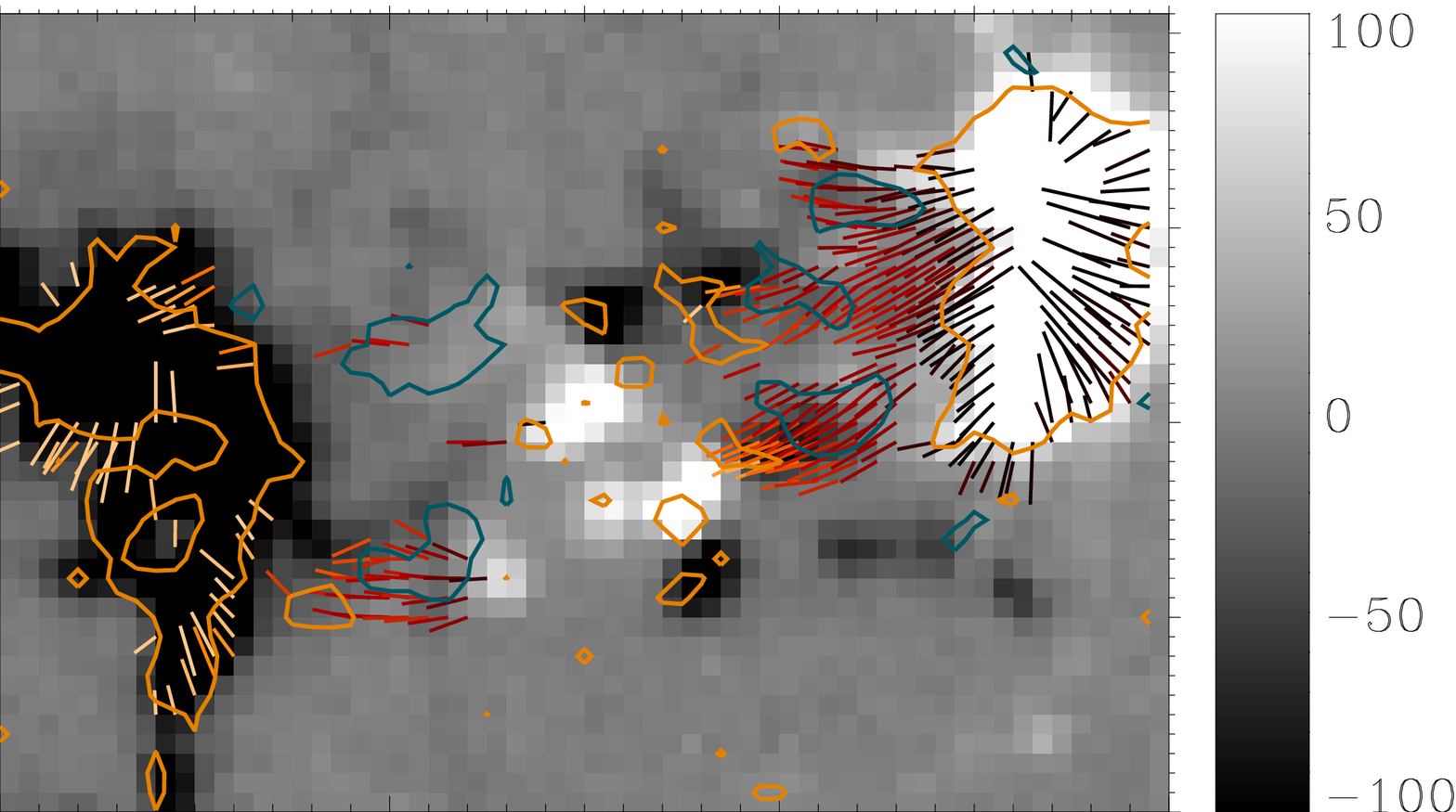}
\caption{Three non-consecutive snapshots in the evolution of AR 11211. The contours
  show the Doppler velocities at -200 (blue, upflows) and 200 (orange,
  downflows) ${\rm ms^{-1}}$ for pixels where the magnetic field
  strength is above the noise level. The background image shows the
  longitudinal magnetic flux density
  saturated at $\pm 100$ gauss (white/black represents positive/negative polarity) and the headless arrows show the
  mostly transverse component of the field. The FOV of these images is
  30 x 20 arcsecs.\label{fig:vcontours}}
\end{center}
\end{figure}

\begin{figure}
\begin{center}
\includegraphics[angle=0, scale=0.40]{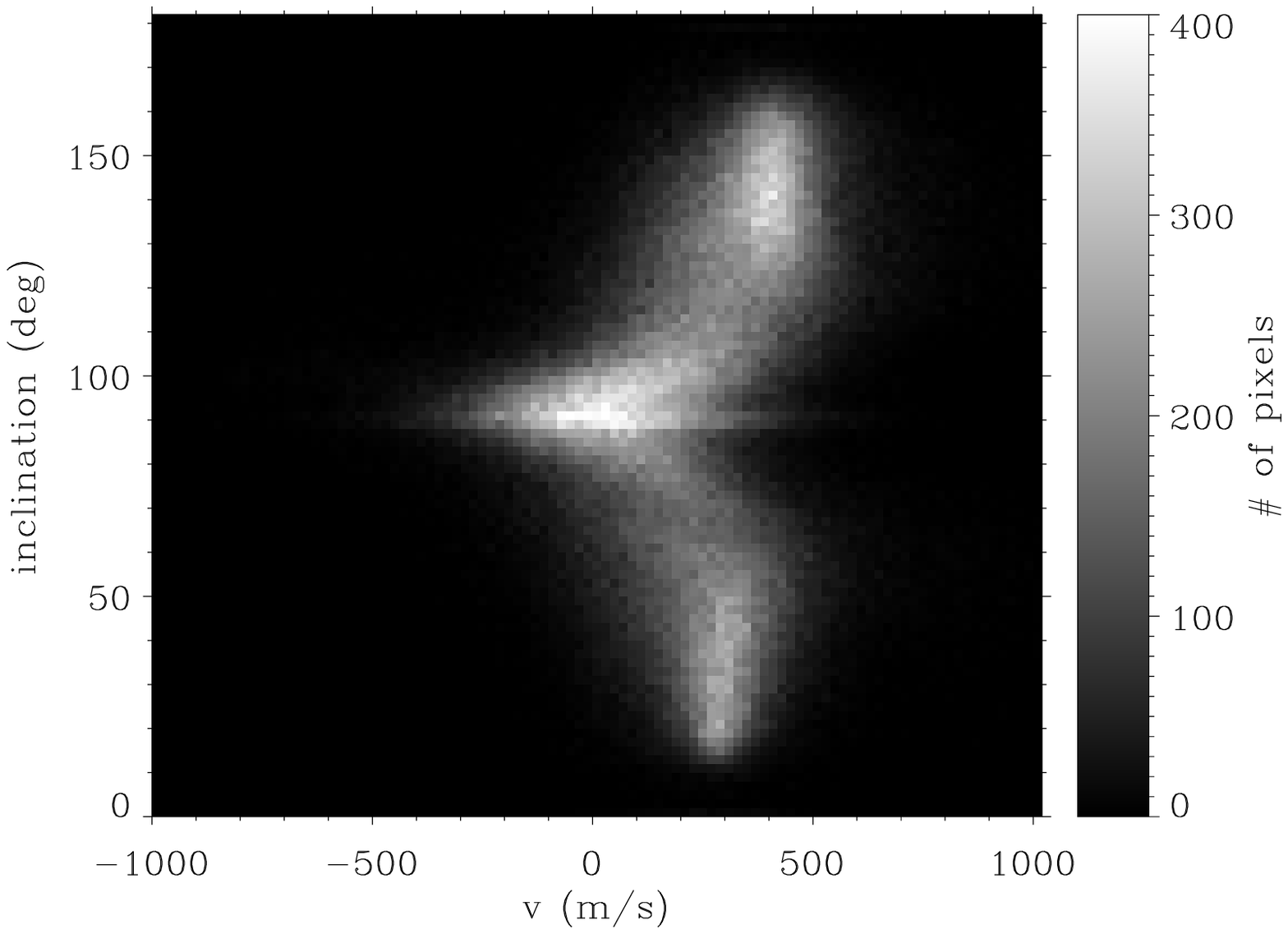}
\includegraphics[angle=0, scale=0.40]{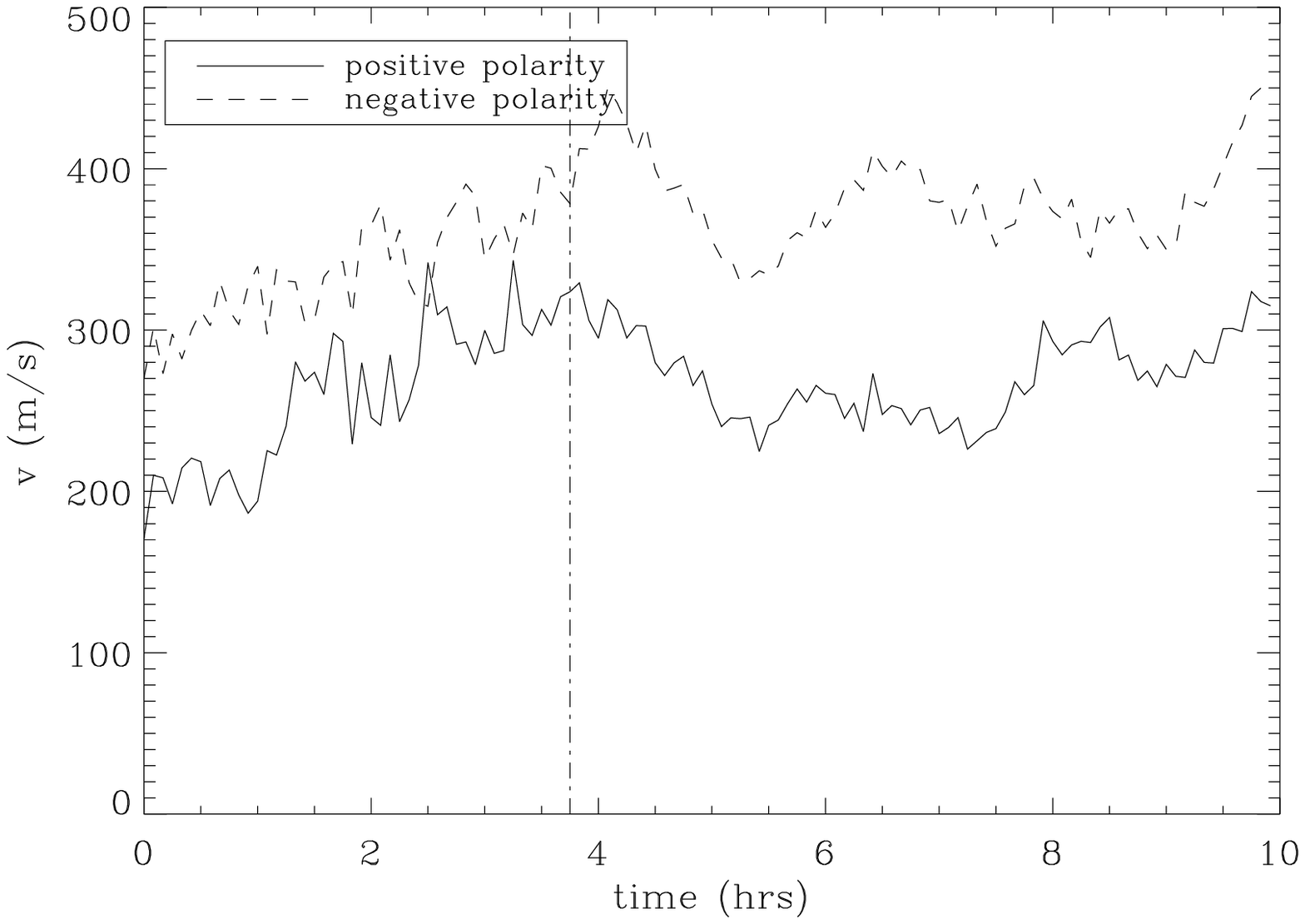}
\includegraphics[angle=0, scale=0.40]{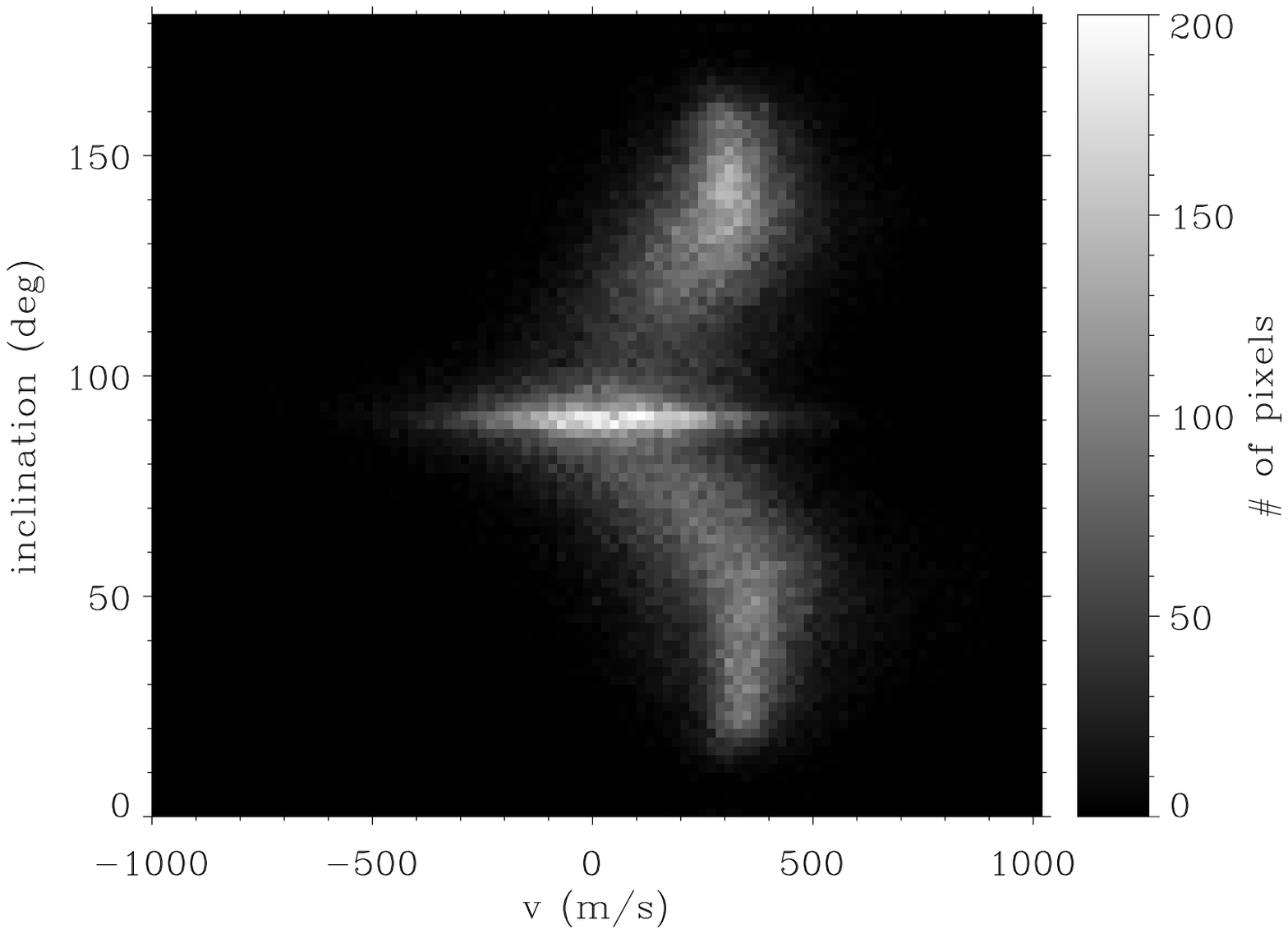}
\includegraphics[angle=0, scale=0.40]{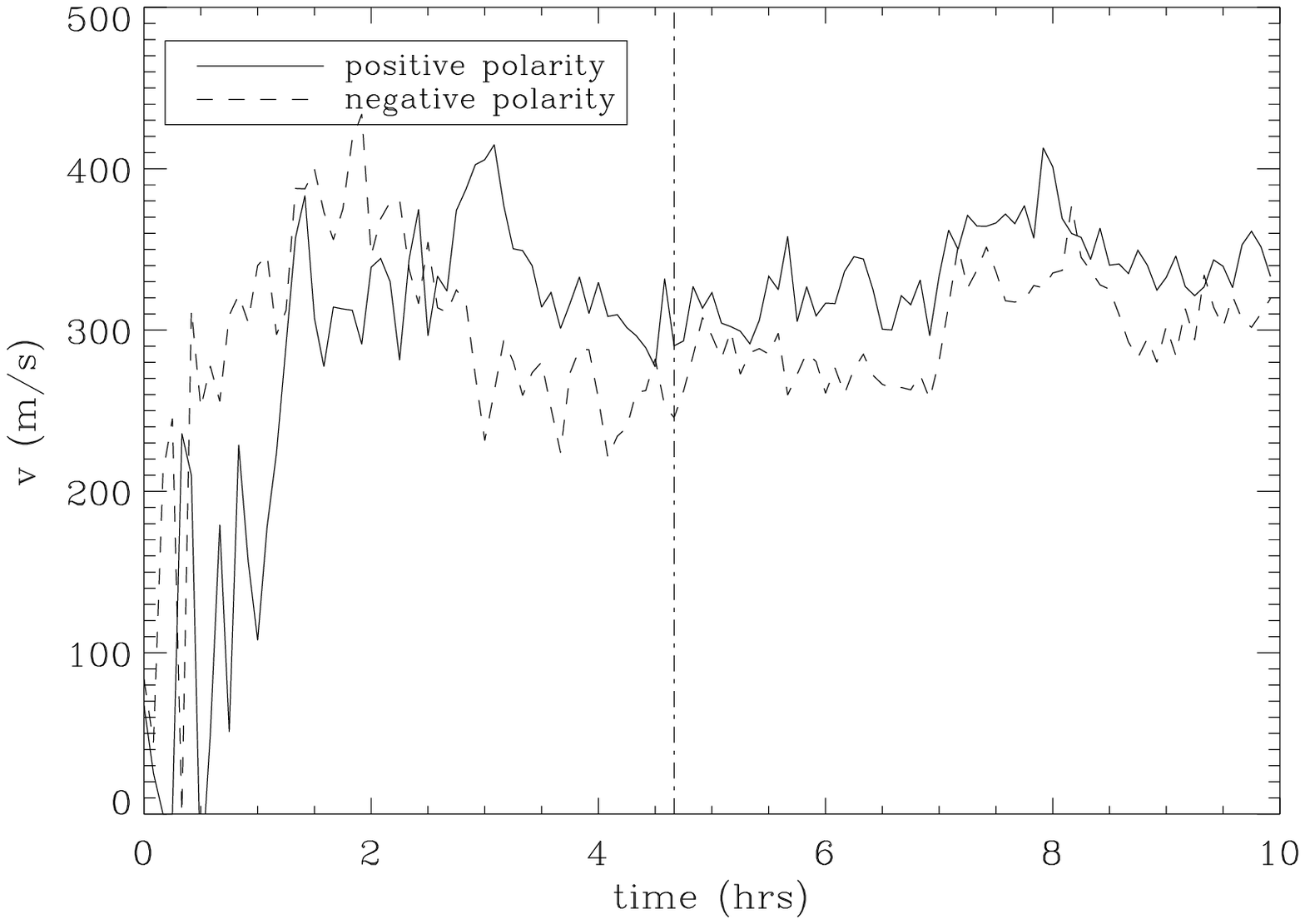}
\caption{Kinematic aspects for AR 11105 (top) and 11211 (bottom). Left: density scatter plots of the magnetic field inclination versus
 the Doppler velocity for both active regions. Each plot combines the
 first 24 hours of emergence and only pixels with magnetic field
 strength above the noise are considered. An inclination of $90^{\circ}$
corresponds to a magnetic field almost parallel to the solar
  surface and negative velocities represent upflows. Vertical fields
  always harbour downflows whilst horizontal fields have a strong
  correspondence with upflowing material. 
Right: temporal evolution of the average downflowing velocity at the
footpoints of both ARs (defined as where the magnetic field is
inclined less than $40^{\circ}$ with respect to the vertical). The
velocities are shown separately for the positive and the negative
polarities. The vertical dash-dotted line marks, approximately, the time at which the
center of the AR passes the central meridian. \label{fig:scatterplots}}
\end{center}
\end{figure}
One of the main characteristics of the emergence process of ARs is the
relation between the plasma flows and the magnetic topology. AR
emergence sites are characterized by downflows of up to $2 \,{\rm
  kms^{-1}}$ in the vicinity of rapidly growing pores and by upflows around the main polarity
inversion line,
where the new magnetic flux is surfacing \citep[see, for instance,][]{1985zwaan, 1985brants_b}.

Figure \ref{fig:vcontours} shows three snapshots in the evolution
of AR 11211. Upflows (blue contours) and downflows (orange contours)
for the magnetized pixels are plotted over the grayscale background
that represents the longitudinal magnetic flux density saturated at
$\pm 100$\,G). Red arrows show the direction of the transverse component of the
magnetic field vector where it is rather inclined with respect to the
LOS (forming an angle of less than $40^{\circ}$ from the solar surface).
In magnetized areas, there is a strong correlation between the
velocity patterns and the magnetic field orientation. Upflows are
present wherever the magnetic field has a predominantly transverse
nature, whilst downflows are systematically correlated to the more
vertical configurations.
Although only the velocity contours for magnetized areas are shown
in figure \ref{fig:vcontours}, it
is important to note that the sizes and velocities of the
upflow patches do not differ significantly from those of the
surrounding quiet Sun at the spatial resolution of the HMI
data. Horizontal field patches brought up to the
surface span spatial scales of several granules (5 to 10 \arcsec).

The left column of figure \ref{fig:scatterplots} shows density scatterplots of the
magnetic field inclination versus the Doppler velocity for the first
24 hours of emergence of each of the active regions. An inclination
of $90^{\circ}$ corresponds to a magnetic field parallel to the solar
surface. It is clear that, during these first stages of the emergence
process, the more vertical fields (with inclinations
close to 0 and $180^{\circ}$) always harbour downflows, whilst the
horizontal component of these fields tends to be accompanied by
upflowing material.

Both scatter plots present a slight asymmetry in the Doppler shifts of
the AR footpoints. In both cases, the leading polarity ($\theta\sim
180^{\circ}$ for AR 11105 and $\theta\sim 0^{\circ}$ for
AR 11211) shows stronger downflows than the
following one - this is especially obvious in the case of AR 11105
(top left panel). This is partly due to the proper motions of the footpoints, which drift apart
as the AR grows. These proper motions have a line-of-sight
component that adds up to the measured Doppler velocity whenever
the AR is not at disk center. Our targets are never very far away from
disk center for the time span of the observations presented in this
paper. However, during the very early stages of emergence (for the first 10-12
hours of the sequence) both ARs are
approaching the central meridian from the East side, whilst, for the second half 
of the sequence, they travel Westwards away from it. 
One would expect to see a larger LOS component of the proper motions of
the footpoints in the second half of the sequence, when the ARs are larger and more
developed. On the West side of the meridian, the projection of the
proper motion of the footpoints is always in the sense that the
leading polarity would seem to be moving away from the observer (redshift) whilst the following
polarity would appear to do the opposite (blueshift). Hence, this could
explain the Doppler shift asymmetry seen in the scatter plots.

The panels on the right column of Figure \ref{fig:scatterplots} show
the temporal evolution of the average velocity at the footpoints of both active regions,
separated by polarity. The vertical dash-dotted line marks the
approximate time when the
center of the AR crosses the central meridian. 
Although AR 11211 (bottom
right panel) presents a slight
asymmetry of the mean downflows at its footpoints, this is relatively
small when compared to the spread in velocity values. Also note that, for a fraction of the time, the
following polarity has stronger downflows than the leading one, but the trend switches when
the AR is about to cross the meridian. This is likely to be an
effect of the footpoint proper motions.

\noindent AR 11105, on the other hand, shows a systematic strong asymmetry that does not switch its behavior
when crossing from the East to the West side of the meridian. The
leading polarity has, on average, $100{\rm
  ms^{-1}}$ stronger downflows than the following one, suggesting that
the imbalance is real and not just a product of transverse motions and
projection effects. It is also interesting to note that there is a
strong correlation between the flows at both footpoints (see how the
solid and dashed lines of the top right panel of figure
\ref{fig:scatterplots} follow each other's trend). This supports the
idea of a net flow going from the following to the leading polarity
(the flow becomes stronger or weaker at both
footpoints in synchrony).

In both ARs, downflows are present in the entire area of the pores,
being stronger at the edges. MDF's, albeit usually less vertical than
the main footpoints of the AR, also harbour consistent downflows
during their short lifetimes.

\subsection{Continuum intensity}

\begin{figure}
\begin{center}
\includegraphics[angle=0, scale=0.7]{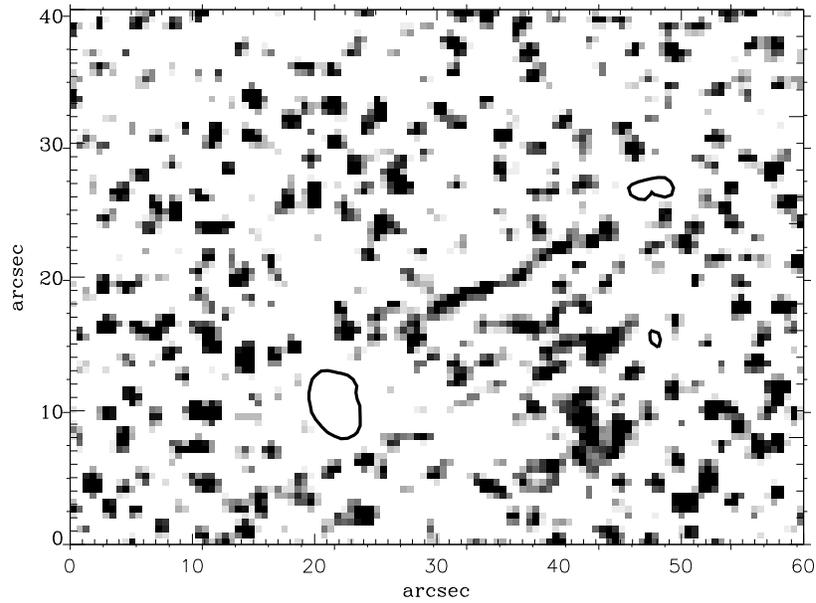}
\caption{Negative snapshot of the continuum intensity of AR 11105, saturated below
  $1.013 \cdot {\rm I_{CONT}}$ (white). Black contours mark the locations of
  the pores over the grayscale background, where only the
  brighter granules stand out. The granulation pattern in between the footpoints
  of the AR often shows fibrilar structures, lower contrasts and
  higher-than-average intenisties.\label{fig:fibrils}}
\end{center}
\end{figure}

The regions of emergence of the top of the magnetic arcades are
characterized by horizontal fields and upflowing velocities. The granulation in these areas is
often elongated in the direction that connects the main footpoints of
the AR \citep[see, for instance][]{schliche2010, cheung2010, stein2011}. Small-scale short-lived dark features
in the Photosphere and the Chromosphere, often followed by brightenings,
accompany the emergence. These events are thought to be the signature
of arch filament systems of the scale of several granules, rising up
from the photosphere to the chromosphere \citep{vargas2012}.

The spatial resolution of the HMI data imposes some constraints on the analysis
of granular-scale events. Nevertheless, the data allow us to witness some of the characteristical
properties of granulation at the emergence site (in between the
main footpoints of the AR). Elongated
bright features are a common sighting in these areas.
Figure \ref{fig:fibrils} enhances these occurrences by
masking out anything that is darker than 1.013 times the average
continuum intensity of the surrounding quiet Sun, highlighting the brighter
granules only (note that it is a negative image, so a lighter shade of gray
represents a lower continuum intensity). The black contours show the
locations of the pores. Elongated bright features can be seen in
between the AR footpoints. They are always are co-spatial with patches
of horizontal magnetic field and parallel to the direction of the
field lines (note
that these are not always necessarily aligned with the general orientation of
the active region, though). Sometimes, short-lived (30 minutes or less) dark filamentary structures also pop up in
these areas, often delineating the brighter elongated ones.

\noindent Emergence sites are also characterized by other granulation
peculiarities. The average continuum intensity in the area in between the
footpoints of the AR exceeds by 0.3-2\% that of the
quiet Sun in both of our datasets. The strongest excesses are
co-temporal and co-spatial with the occurrance of the strongest
patches of horizontal fields. Also, despite the brightenings and darkenings taking place, 
granulation has a systematically lower rms contrast in
the emergence site than in the quieter places in the vicinity of the AR.
This excess intensity of the granulation in the emergence
zone could be a consequence of the larger radiative losses of the hotter plasma that is dragged
up to the surface by the buoyant fields. It could also be
 due to the pressure balance requirement in and around the
emerging site. The weak horizontal fields (200-500 G) create a
magnetic pressure that has to be compensated by a lower gas pressure. This leads to a less dense
environment that is consequently more transparent so the spectral lines will
form at deeper (hotter) layers.

\section{Discussion}\label{sec:discussion}

This paper focuses on the ``naked'' emergence of active
regions from their first imprints on the solar surface. 
With this purpose in mind, we follow the first 24 hours of existence of two
relatively-isolated active regions using sequences of photospheric vector magnetic fields and Doppler
velocities obtained with the Helioseismic and Magnetic Imager on board SDO. 

AR 11105 grows faster and at 3 times the flux rate of AR
11211. It brings intrinsically stronger transverse fields to the
surface (up to 1000\,G as opposed to 400\,G)  and harbours stronger
vertical fields at its footpoints. It also presents a systematic
asymmetry in the downflowing plasma at its footpoints, with the leading
polarity having, on average, $100{\rm ms^{-1}}$ faster downflows
than the following one. This characteristic together with the strong
correlation in the strength of the flows at both footpoints, suggests the existence of a net flow from
the following to the leading polarity, in agreement with the findings
by \citet{cauzzi1996} and in disagreement with certain models of flux
emergence \citep{fan1993}. This flow asymmetry is not
clear in the smaller AR.

\begin{figure}
\begin{center}
\includegraphics[angle=0, scale=0.15]{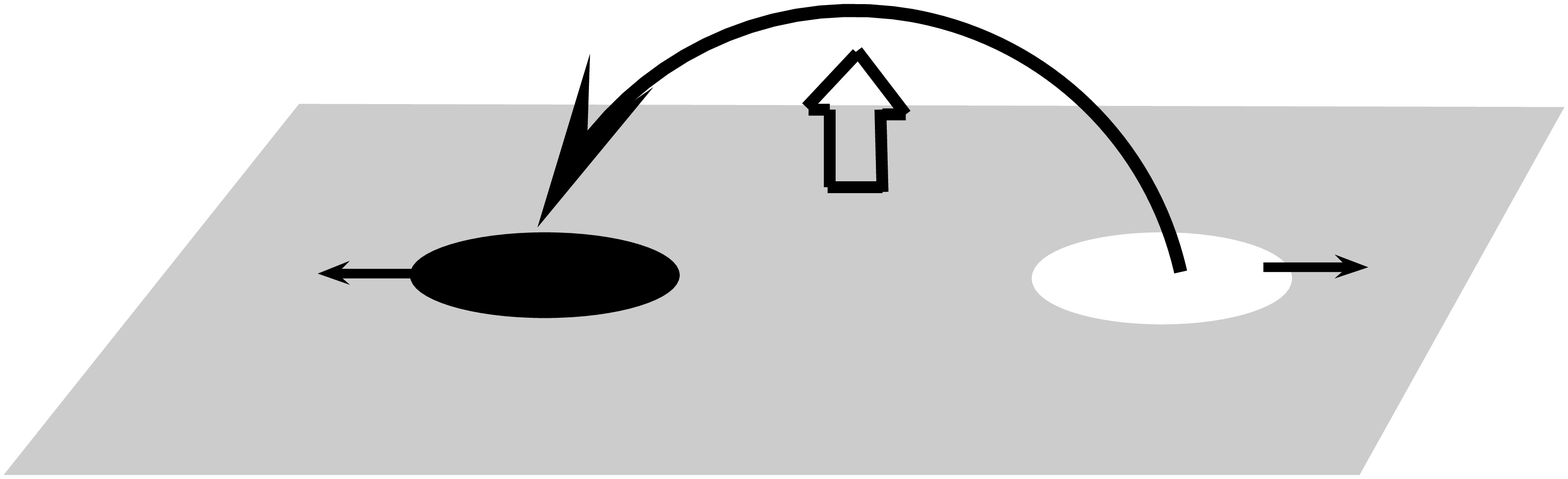}
\includegraphics[angle=0, scale=0.15]{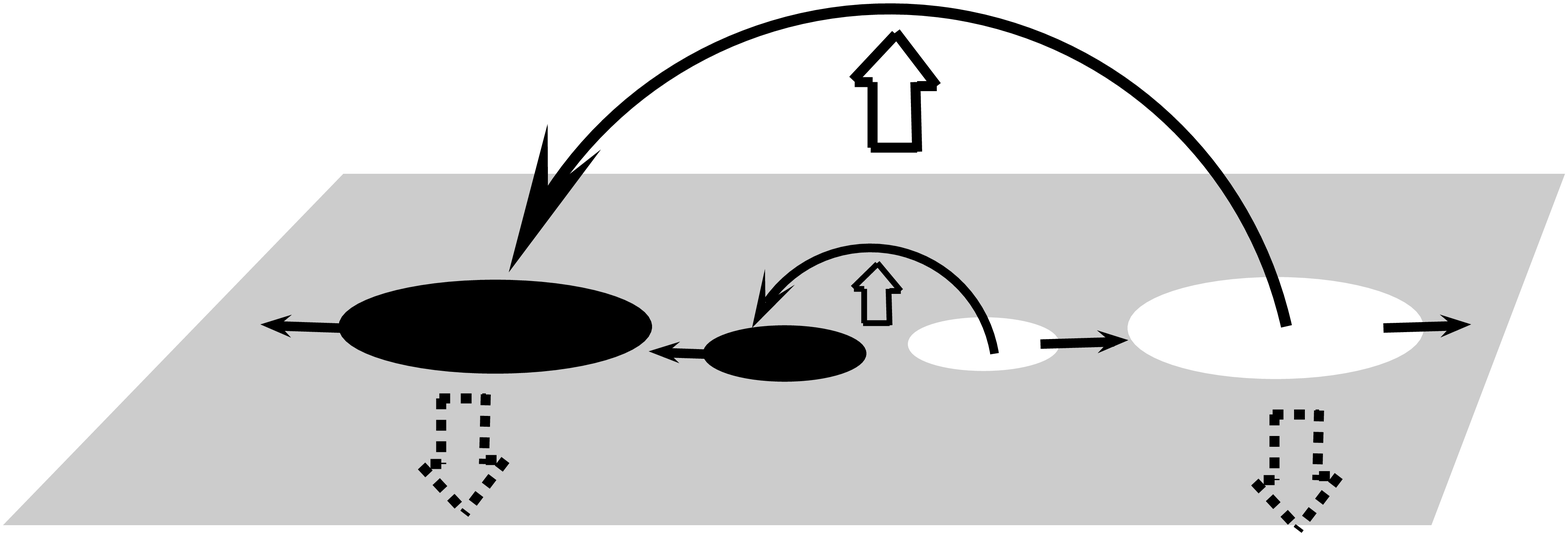}\\
\includegraphics[angle=0, scale=0.2]{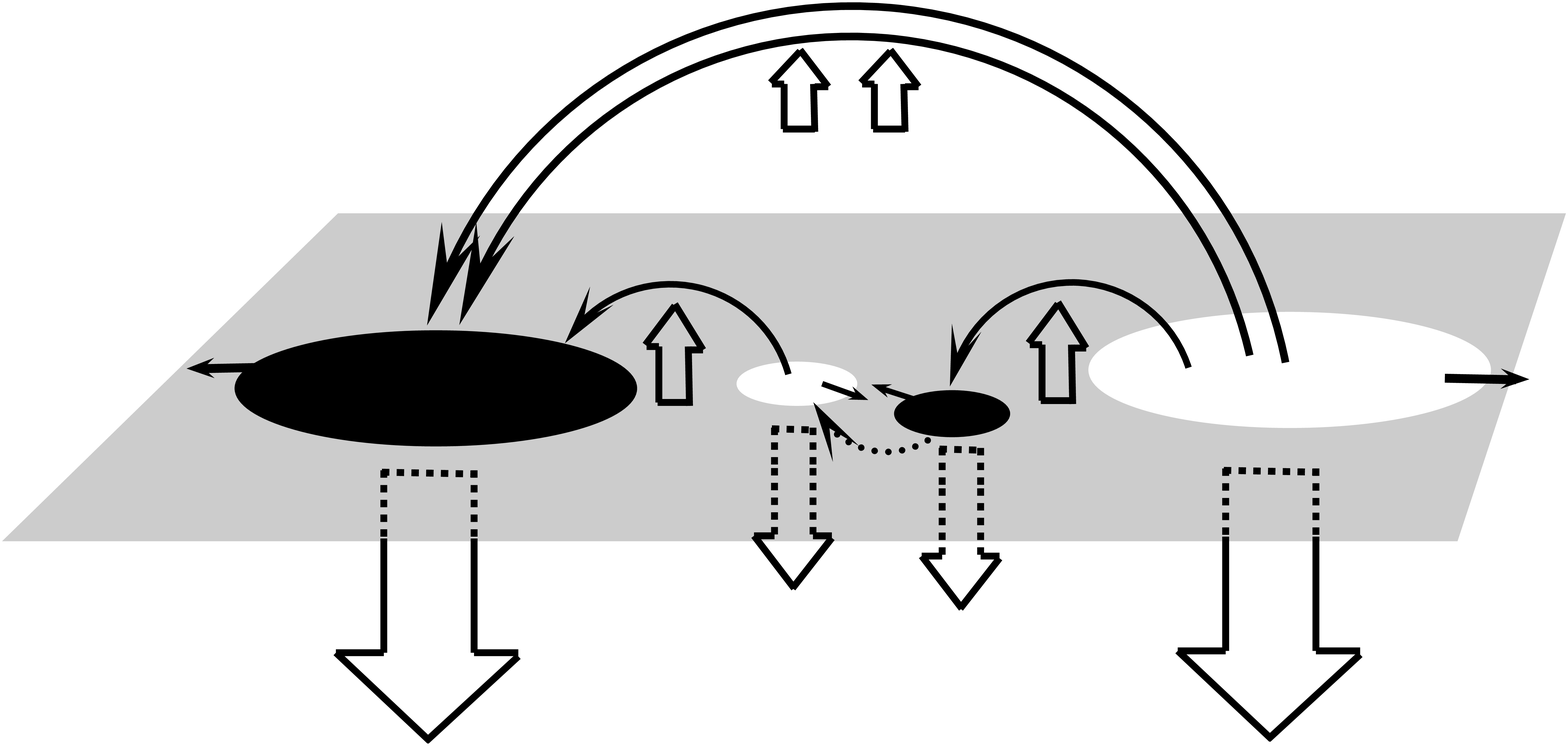}
\caption{Sequence of the naked emergence of an AR. The first photospheric
  signature is a small dipole connected by horizontal fields. Patches
  of transverse fields continue to emerge with their corresponding
  footpoints, which drift and merge with the pre-existing ones. When
  the AR is large enough, the magnetic arcade is not able to lift the
  weight of the plasma that it carries and remains trapped to the
  photosphere at the locations of the MDFs. The poles of the MDFs
  approach one another and cancel out, liberating the field lines and
  letting the arcade rise into the corona.\label{fig:cartoon}}
\end{center}
\end{figure}

Qualitatively, however, both ARs display a
similar behaviour during emergence.
The process of the ``naked'' emergence of an AR is simplified in the
cartoon of figure \ref{fig:cartoon}.
The very first signature of an AR on the surface is a relatively
simple dipole connected by horizontal fields. Small (5 - 15\arcsec) bursts 
of transverse field patches flanked by more vertical footpoints keep bringing new
flux to the surface. As the vertical magnetic features coalesce, the AR starts growing
appreciably and its main footpoints differentiate themselves and drift apart. Systematic
downflows begin to dominate at the places where the field is anchored
to the photosphere. Once the AR reaches a certain size, the patches of
horizontal field do not span the entire distance between the main footpoints
anymore. Instead, they connect via an intermediate layer of MDFs (dipoles of 
opposite polarity to that of the AR) that appears in between the two
polarities.
 Upflows characterize the patches of transverse field while
plasma drains back down into the photosphere through all of the
places where the field is pinned down to the surface. 
Brighter granulation, elongated in the direction of the magnetic
field, often accompanies the horizontal patches. This characteristic
is more evident the stronger the magnetic flux is.
When the AR is big enough, several layers of MDFs coexist in
the emergence zone. The
magnetic field contributing to the global emergence serpentines into and out of the surface from the main
positive polarity to the negative one, process that is commonly
referred to in the literature as {\em resistive emergence}
\citep[e.g.][]{pariat2004, pariat2009}. In this scenario, the MDFs are just the drainage points of
the plasma carried by the upwelling field lines. The poles of MDFs
tend to approach each other and cancel out, presumably reconnecting 
somewhere close to the surface (below or above) and releasing the field lines that are
now free to continue rising up into the corona. This is consistent with
 the scenario described by \citet{cheung2010}, where the cancelling MDFs are the
visual signature of the reconnection of field lines and the discharge of
mass from the rising magnetic structure.

\section{Acknowledgments}
The National Center for Atmospheric Research is sponsored by the
National Science Foundation. This manuscript was mostly written while RC was
a visiting scientist at the Kiepenheuer-Insitut f\"ur Sonnenphysik
(Freiburg, Germany). RC wishes to thank Rolf Schlichenmaier, Juan
Borrero, Wolfgang Schmidt, Valent\'\i n Mart\'\i nez Pillet and Alfred
de Wijn for insightful comments on this piece of research.


\end{document}